\documentclass[%
 jmp,
 amsmath,amssymb,
preprint,%
]{revtex4-1}

\usepackage{graphicx}
\usepackage{dcolumn}
\usepackage{bm}

\usepackage[utf8]{inputenc}
\usepackage[T1]{fontenc}
\usepackage{mathptmx}
\usepackage{etoolbox}
\usepackage{amsfonts,amssymb,amsmath,graphicx}
\usepackage{amsthm}

\makeatletter
\def\@email#1#2{%
 \endgroup
 \patchcmd{\titleblock@produce}
  {\frontmatter@RRAPformat}
  {\frontmatter@RRAPformat{\produce@RRAP{*#1\href{mailto:#2}{#2}}}\frontmatter@RRAPformat}
  {}{}
}%
\makeatother

\newtheorem{theorem}{Theorem}[section]
\newtheorem{lemma}[theorem]{Lemma}

\theoremstyle{definition}
\newtheorem{definition}[theorem]{Definition}

\theoremstyle{remark}
\newtheorem{remark}[theorem]{Remark}

\numberwithin{equation}{section}

\newtheorem{proposition}{Proposition}
\newtheorem{corollary}{Corollary}

\newcommand{\dv}{{\rm d}}


\begin{document}

\preprint{AIP/123-QED}

\title{Beyond mean field: condensate coupled with pair excitations}
\author{Stephen Sorokanich}
 \affiliation{Department of Mathematics, University of Maryland, College Park.}

 \email{ssorokan@umd.edu}

\date{\today}

\begin{abstract}
We prove existence results for a system of partial differential equations describing the approximate condensate wavefunction and pair-excitation kernel of a dilute (T=0) Bose gas in the stationary setting, in the presence of a trapping potential and repulsive pairwise atomic interactions. Notably, the Hartree-type equation for the condensate in this system contains contributions from non-condensate particles, and the pair excitation kernel satisfies a nonlinear operator equation. The equations studied here are inspired by the work of Griffin, who derived this system in the study of finite temperature condensates. The techniques employed include a variational principle, which exploits the connection between unitary Bogoliubov rotations and a nonlinear operator equation for the pair excitation kernel. An iterative procedure for constructing solutions is also included. 
\end{abstract}

\maketitle

\section{Introduction}
\label{sec:Intro}

In this paper, we study a nonlinear system of integro-partial-differential equations which arise in the study of a trapped Bose-Einstein condensate (BEC) in equilibrium with particles outside the condensate. The equations are derived (in a physically-inspired although heuristic fashion) from quantum mechanical first principles via a model for a finite-but-large number of interacting bosons, inspired by several preeminent works in physics \cite{Griffin1996, fetter72,wu61}. The quantities described by this system are the macroscopic condensate wavefunction, $\phi(x):\mathbb{R}^3 \to \mathbb{C}$, and the pair-excitation kernel, $k(x,y):\mathbb{R}^{3\times 3}\to \mathbb{C}$. Together, these quantities provide a description of the quantum ground state of a dilute Bose gas which generalizes the mean-field theory \cite{MGM, GrillakisMargetisSorokanich,Margetis2012, GMM2010}. 

The mean-field theory for BEC specifies a nonlinear Schr{\"o}dinger equation for the single-particle condensate wavefunction, $\phi(x)$. Assuming that the effective ground state of the system is a tensor product involving only this mean-field condensate function yields an approximation to the exact many-body Hamiltonian which is \textit{quadratic} and \textit{diagonal} in field operators for states orthogonal to the condensate. This is not true for models that account for the effects of pair-excitation. Such models can be described by two functions, referred to as $\phi(x)$ and $k(x,y)$ in this work, and are associated with an approximate Hamiltonian that is quadratic, but not necessarily diagonal, in the field operators. The heuristic approximation of the exact many-body Hamiltonian considered in this work takes includes contributions from all two-point correlation functions associated with  non-condensate particles.

The models and techniques presented here, as well as the use of fundamental quantities $\phi,\,k$ to describe the pair-excited ground state, are inspired by several well-known works in the physical modeling of Bose-Einstein condensation. The quantum field theory of BEC originated in the work of Bogoliubov, Gross, and Pitaevskii \cite{Bogoliubov1947, Gross61, Pitaevskii61}, who formulated the mean-field approximation and the concept of macroscopic wavefunction for the condensate (proportional to $\phi$ in our terminology). Lee, Huang, Yang, and Luttinger began a comprehensive study of the Bose gas in the periodic box in the 1950s \cite{leehuangyang,leeyang,HuangYang1957,huangyangluttinger} and derived some of the first corrections to the ground state energy predicted by the mean field theory; these corrections are produced by correlations between the condensate and pairs of non-condensate particles with equal-and-opposite momenta. This early work for the periodic box was adapted to the non-periodic setting in the 1960s by T.~T.~Wu \cite{wu58,wu61}, who was the first to formulate the pair-excitation kernel (the quantity $k$), and derived the fundamental correction to the mean-field ground state represented schematically by:
\begin{equation}\label{WuSchematic} \Psi_0 = \exp(P[k])\Psi_{\mathrm{mf}}[\phi].
\end{equation}
Here $\Psi_\mathrm{mf}[\phi]$ is the tensor-product ground state of the mean-field theory, and $P[k]$ is the (non-Hermitian) pair-excitation operator.

The primary motivations for this work are the quadratic Hamiltonian models of Fetter (1971) \cite{fetter72}, and Griffin (1996) \cite{Griffin1996, GriffinBook}. These authors formulate the unitary (Bogoliubov) rotation of quadratic, non-translation-invariant Hamiltonians describing systems in the presence of a trapping potential. By proper choice of this unitary transformation (which amounts to the choice of a non-orthogonal basis of single-particle states), a diagonal transformed Hamiltonian is obtained, and the excitation spectrum can be directly inferred. The operator basis in which the Hamiltonian is diagonalized is commonly referred to as the basis of quasiparticles \cite{huang2008statistical}. We instead approach the mentioned quadratic models from the formalism of pair-excitation. In previous work \cite{GrillakisMargetisSorokanich, Sorokanich-Thesis}, we showed that there is an intimate connection between the quasiparticle basis which appears in the work of Fetter \cite{fetter72}, and the pair-excitation kernel that appears in the second-order correction to the ground state described by Wu \cite{wu61}, and represented by the expression \eqref{WuSchematic}. The connection between this kernel and the single-particle quasiparticle basis parallels results in operator theory (see, e.g., the works of Tretter \cite{Tretter-book,Tretter2016}, Albeverio, et. al. \cite{AlbeverioMotovilov2010,AlbeverioMotovilov2019, Kostrykin03-chap}). An existence theory for this kernel in the specific context of the Bose gas was developed in \cite{GrillakisMargetisSorokanich}, based on a variational principle.

The purpose of this paper is twofold. First, we apply the method of pair-excitation to a family of quadratic models for the Bose gas which are not described by our previous work. This extension is nontrivial, since the model in this work contains nonlinear couplings between the condensate and excited states (in either the quasiparticle representation or in the representation using the pair-excitation kernel). Second, we utilize the new representation of this system in terms of the excitation kernel to give an existence theory for the resulting time-independent nonlinear equations. The operator equation for $k$ in this model will contain nonlinear terms which represent non-condensate particle densities. We will show how it is possible to solve the nonlinear system for $k$ and $\phi$ using a variational approach. Two methods for solving the variational problem are given; the first method is general and is a direct extension of our previous method. The second method is an iteration scheme, which gives more detailed information about the solution found.

Our primary source in the physics literature for the model studied here is Griffin \cite{Griffin1996, GriffinBook}, although it is sufficiently general to be called the Bogoliubov de Gennes system \cite{Bach2022}. The excitations of a Bose gas in non-translation invariant settings has been studied by many authors; see,
e.g., \cite{Dalfovoetal1999, Gardiner1997, Leggett2001, Ozeri2005, Rovenchak2016}. While time-dependent systems related to the one studied here have been the subject of past work \cite{Bach2022, Nam2017,GM2013-a, GM2013-b}, we believe that our focus on the stationary case can potentially shed new light onto the problem. As far as the author is aware, this is the first rigorous treatment of the explicitly stationary problem, and the first time these equations have been linked to a pair excitation formalism. In addition to these works, we refer the reader to recent efforts to place the quadratic approximation of bosonic Hamiltonians on a rigorous basis \cite{Seiringer,NamNapiorkowski2017, NamNapiorkowski2017-II, NamSolovej2016, NamSeiringer2015, BrenneckeSchlein2019}; the model presented here has potential to be rigorously treated as an approximation to the exact Schr\"{o}dinger dynamics over quasifree states \cite{Zagrebnov2001}, although we do not follow this path in this work.

\subsection{Mathematical prerequisites and Hamiltonian model}
We now introduce the exact Hamiltonian for interacting bosons, and the heuristic reduction of this Hamiltonian to a quadratic expression in field operators inspired by the work of Griffin \cite{Griffin1996,GriffinBook}. The Bogoliubov rotation of the resulting reduced Hamiltonian is introduced, and the mean field equations for the quantities $\big(\phi(x), u_j(x), p_j(x)\big)_{j=1}^\infty$ are derived. We then describe the equivalent description of these quantities in terms of the condensate wavefunction and the pair-excitation kernel, $\big(\phi(x), k(x,y)\big)$. This subsection will also serve to introduce the notation and terminology used throughout the rest of the work.

Our starting point is the bosonic Fock space, denoted $\mathbb{F}$, which is defined as the direct sum of all nn-particle symmetric Hilbert spaces,
$$\mathbb{F} := \mathbb{C}\oplus \bigoplus_{n=1}^\infty{L^2_\mathrm{sym}(\mathbb{R}^{3n})}.$$

The operators $a_x, a_x^\ast$ are the usual bosonic field operators on $\mathbb{F}$, which satisfy the canonical commutation relations:
\begin{equation}
[a_x, a_y] = 0 = [a_x^\ast, a_y^\ast],\quad [a_x, a_y^\ast] = \delta(x-y). \quad x,y\in\mathbb{R}^3.
\end{equation}
Here, as in the rest of the work, $\delta(x-y)$ denotes the Dirac mass of the delta function. The exact many-body Hamiltonian, $\mathcal{H}$, for $N$ bosons with a repulsive, mean-field, pairwise interaction $\upsilon_N$, in the presence of trapping potential $V_\mathrm{trap}(x)$, is given in second-quantized form by the Fock space operator: 
\begin{equation}\label{Ham1}
\mathcal{H} = \int{\dv x\, \{a_x^\ast \big(-\Delta + V_\mathrm{trap}(x)\big)a_x\}}+\frac{1}{2}\iint{\dv x\dv y \,\big\{a_x^\ast a_y^\ast \,\frac{1}{N}\upsilon_N(x-y) a_x a_y\big\}}.
\end{equation}
The Hamiltonian $\mathcal{H}$ includes the mean-field type interaction, $\frac{1}{N}\upsilon_N(x)$, which is required for the analysis that follows. Specifically, we take $\upsilon_N(x) = N^{3\beta}\upsilon(N^\beta x)$, where $\upsilon$ is a smooth, positive, radially symmetric function, $0<\beta\le 1$, with $\int{\dv x\{\upsilon(x)\}} = g$, for $g>0$. As $N\to\infty$, $\upsilon_N$ is assumed to approximate the delta function. The trapping potential $V_\mathrm{trap}(x)$ is assumed to be positive, and moreover, we assume that the single-particle Schr\"{o}dinger operator $(-\Delta_x + V_\mathrm{trap}(x))$ has a positive spectrum with a gap between it's lowest and first excited states. It is useful to take the concrete example $V_\mathrm{trap}(x) = |x|^2$ for $x\in\mathbb{R}^3$ as the prototypical trapping potential. We also introduce the number operator, $\mathcal{N}:\mathbb{F}\to\mathbb{F}$, defined by 
\begin{equation}\label{numberop}
\mathcal{N} := \int{\dv x \{a_x^\ast a_x\}}.
\end{equation}
Since the following approximation scheme will not conserve the total number of particles, we consider instead the operator $\mathcal{K} := \mathcal{H} - \mu\mathcal{N}$, where $\mu>0$ is the chemical potential, a Lagrange multiplier constraining the average total number of particles.

The assumptions on the interaction potential yield the following $L^\infty$ estimate
\begin{equation}\frac{1}{N}\upsilon_N(x-y)\le \|\upsilon\|_{L^\infty},\quad 0\le \beta \le 1/3,\end{equation} 
and by Young's convolution inequality (where the operation $\ast$ denotes convolution), the estimate
\begin{equation}\|\frac{1}{N}(\upsilon_N\ast f)\|_{L^2}\le g\|f\|_{L^2},\quad f\in L^2(\mathbb{R}^3).\end{equation}

The exact operator $\mathcal{K}$, with $\mathcal{H}$ coming from equation \eqref{Ham1}, is a \textit{quartic} expression in field operators $a,\ a^\ast$ due to the presence of the two-particle interaction. The main question that motivates the quadratic reduction of this exact Hamiltonian is: what is the low-lying spectrum of $\mathcal{K}$, and can this spectrum be approximated by the spectrum of an effective \textit{quadratic} operator? We do not take up this question in rigorous detail here; our goal is rather to study a simplified effective model which incorporates features of the many-body system for states close to the ground state, and we take the lead from the significant work in physics already mentioned. We emphasize the utility of an approximation which is quadratic in field operators orthogonal to the condensate, since quadratic Hamiltonians can be diagonalized by unitary rotation. Two systematic, although heuristic, approximation schemes are now summarized, motivated by the works \cite{Griffin1996} and \cite{fetter72} respectively, and the unitary rotation of the resulting quadratic operators. For the sake of completeness, elements of the reduction of operator $\mathcal{K}$ are given. Details of specific calculations are omitted, and we refer to \cite{Griffin1996, GriffinBook, fetter72} for the ultimate justifications that these authors give. The approximation schemes are heuristic, since they do not address, e.g., how accurately the ground state of the reduced Hamiltonian approximates the ground state of Hamiltonian \eqref{Ham1}.

\textbf{Formal reduction (Griffin, 1996 \cite{Griffin1996}):} 
First, the heuristic \textit{Bogoliubov approximation} is employed by which the field operators $a_x^\ast,\,a_x$ appearing in the operator $\mathcal{K} = \mathcal{H} - \mu\mathcal{N}$ are decomposed into a macroscopic scalar field proportional to the condensate wavefunction $\phi(x)$, in addition to field operators, $b^\ast_x,\,b_x$, which create/annihilate particles in states orthogonal to the condensate, viz.,
\begin{equation}\label{Bogapprox}
a_x \approx \sqrt{N}\,\,\overline{\phi(x)} + b_x,\quad 
a_x^\ast \approx \sqrt{N}\phi(x) + b^\ast_x.
\end{equation}
The symbol ``$\approx$" is used here, and throughout this section, to denote the fact that \eqref{Bogapprox} is a physically-motivated, but ultimately formal substitution in the current treatment (referring, e.g., to the references \cite{Seiringer, Zagrebnov2001, NamNapiorkowski2017, NamNapiorkowski2017-II} for a nuanced treatment of this canonical approximation). The operators $b_x,\, b_x^\ast:\mathbb{F}\to\mathbb{F}$ are understood to satisfy the following commutation relations:
\begin{equation}\begin{split}\label{bcommutation}
[b_x^\ast,b_y^\ast] &= 0 = [b_x, b_y], \\
[b_x, \,b^\ast_y] &= \delta(x-y) - \phi(x)\overline{\phi(y)}:= \widehat\delta(x-y),\quad x,y\in\mathbb{R}^3.
\end{split}\end{equation}

After making the Bogoliubov approximation, operator $\mathcal{K}$ is further reduced by replacing cubic and quartic terms in $b_x,b_x^\ast$ with effective quadratic operators. This replacement depends on the three (as yet unspecified) functions, which we denote $m^\mathrm{pair}(x,y),\,n^\mathrm{pair}(x,y),\,\rho^\mathrm{pair}(x)$, via,
\begin{equation}\begin{split}\label{quartic}
b_{x}^\ast b_{y}^\ast b_{x}b_{y} &\approx \,\,\,\,2n^\mathrm{pair}(x,y) b^\ast_{y}b_{x} + 2\rho^\mathrm{pair}(x) b^\ast_{y}b_{y} + \overline{m^\mathrm{pair}(x,y)} b_{x}b_{y} +  m^\mathrm{pair}(x,y) b^\ast_{x}b^\ast_{y}\,\,, \\ 
b^\ast_{x} b_{x} b_{y} &\approx \,\,\rho^\mathrm{pair}(x) b_{y} + n^\mathrm{pair}(x,y) b_{x} + m^\mathrm{pair}(x,y) b^\ast_{x}\,\,.
\end{split}
\end{equation}
The specific combinations which appear in \eqref{quartic} are consistent with their role in \cite{Griffin1996}, and will soon be clarified. With the substitutions \eqref{Bogapprox} and \eqref{quartic}, operator $\mathcal{K}$ becomes a quadratic expression involving the undetermined quantities $\phi(x),\, m^\mathrm{pair}(x,y),\, n^\mathrm{pair}(x,y),\,\rho^\mathrm{pair}(x)$, which we rewrite as the following `approximate' Hamiltonian:
\begin{equation}\label{HGrif}\begin{split}
&\mathcal{H}_\mathrm{approx} := \mathrm{const} + \int{dx \big\{\overline{B(x)} b_x + B(x) b_x^\ast\big\}} \\
&+\int{dx dy\Big\{(b_x, b_x^\ast)\begin{pmatrix}
 -H^T(x,y) & ( \upsilon_N \overline{m})(x,y) \\
 -( \upsilon_N m)(x,y) & H(x,y)
\end{pmatrix} \begin{pmatrix}
 -b_y^\ast \\ b_y
\end{pmatrix}\Big\}},
\end{split}\end{equation}
with the operators $H(x,y)$, $B(x)$ in this expression defined by
\begin{equation}\label{8.5}\begin{split}
H(x,y) &:= \{-\Delta_x + V_\mathrm{trap}(x) + (\upsilon_N\ast\rho^\mathrm{pair})(x)-\mu\}\delta(x-y)+ (\upsilon_N n)(x,y), \\
B(x) &:= \big(-\Delta_x + V_\mathrm{trap}(x) + \frac{1}{N}(\upsilon_N\ast \rho)(x) - \mu\big)\phi(x) \\
&+ \int{dy\{(\upsilon_N n)(x,y)\phi(y)+(\upsilon_N m)(x,y)\overline{\phi(y)}\}},
\end{split}
\end{equation}
for,
\begin{equation}\begin{split}
n(x,y) &:= \phi(x)\overline{\phi(y)} + \frac{1}{N}n^\mathrm{pair}(x,y),\quad (\upsilon_N n)(x,y):=\upsilon_N(x-y)n(x,y)\\
\quad m(x,y) &:= \phi(x)\phi(y) + \frac{1}{N}m^\mathrm{pair}(x,y), \quad (\upsilon_N m)(x,y):=\upsilon_N(x-y)m(x,y)\\ 
\rho(x) &:= n(x,x), \quad \rho^\mathrm{pair}(x) := n^\mathrm{pair}(x,x).\end{split}
\end{equation}
Observe that both the operator kernels $B(x)$ and $H(x,y)$ contain the single-particle Schr{\"o}dinger operator $\big(-\Delta_x +V_\mathrm{trap}(x)-\mu\big)$, along with additional terms related to the density of particles in the condensate, as well as $n^\mathrm{pair},\,m^\mathrm{pair},\,\rho^\mathrm{pair}$, which, following Griffin, are identified with marginal densities of particles outside the condensate (equation \eqref{marginal} below).
The Hartree-type equation for the mean-field condensate $\phi(x)$ results by enforcing the condition $B(x) \equiv 0$, 
\begin{equation}\label{Griffinphi}\begin{split}
\mu\phi(x) &= (-\Delta_x + V_\mathrm{trap}(x))\phi(x) + (\upsilon_N\ast |\phi|^2)(x)\phi(x) + \frac{1}{N}(\upsilon_N\ast \rho^\mathrm{pair})(x)\phi(x) \\
&+ \frac{1}{N}\int{dy\{\upsilon_N(x-y)n^\mathrm{pair}(x,y)\phi(y)\}} + \frac{1}{N}\int{dy\{\upsilon_N(x-y)m^\mathrm{pair}(x,y)\overline{\phi(y)}\}}.
\end{split}\end{equation}
The mean field condensate $\phi$ thus depends on the quantities $n^\mathrm{pair},\,m^\mathrm{pair},\,\rho^\mathrm{pair}$. Griffin proposes a closure to this system, i.e., a determination of all functions $\{\phi,\,m^\mathrm{pair},n^\mathrm{pair},\rho^\mathrm{pair}\}$, via the hypothesis that these functions are ground-state correlations in the second-order approximation to $\mathcal{K}$. We write these correlations using bra-ket notation:
\begin{equation}\label{marginal}
\langle b_x^\ast b_y\rangle := n^\mathrm{pair}(x,y),\quad \langle b_x b_y\rangle := m^\mathrm{pair}(x,y),\quad \langle b_x^\ast b_x\rangle :=\rho^\mathrm{pair}(x).\end{equation}
Under this assumption, diagonalization of the effective operator provides the algebraic closure of the system. We now briefly explain how this is done by introducing the Fock space \textit{quasiparticle operator basis} $\alpha_j,\,\alpha_j^\ast:\mathbb{F}\to \mathbb{F}$, for $j\in\mathbb{N}$ and a basis of single-particle states $\{u_j(x),\,p_j(x)\}_{j=1}^\infty$ for the Hilbert space orthogonal to the condensate, denoted $(\phi)_\perp$. These quantities are defined so that the quasiparticle creation and annihilation operators $\alpha^\ast_j,\,\alpha_j$ satisfy the canonical commutation relations
$$[\alpha_j,\,\alpha_k^\ast] = \delta_{jk},\quad [\alpha_j,\alpha_k]=[\alpha_j^\ast,\alpha_k^\ast] = 0,$$ 
and such that the field operators $b_x,b_x^\ast$ have the following decomposition
\begin{equation}\label{bformula}\begin{split}
b_{x} = \sum_j{\{u_j(x)\alpha_j - \overline{p_j(x)}\alpha_j^\ast\}},\quad
b^\ast_{x} = \sum_j{\{\overline{u_j(x)}\alpha_j^\ast - p_j(x)\alpha_j\}}.
\end{split}
\end{equation}
The commutation relations \eqref{bcommutation} for $b_x,b_x^\ast$ imply the following completeness relations for the basis of single-particle states (for all $j,j'\in\mathbb{N}$ and $\delta_{jj'}$ denoting Kronecker's delta):
\begin{equation}\label{eq:UP-relns}
\begin{split}
\int \dv x\ \{u_j(x)p_{j'}(x) - p_j(x)u_{j'}(x)\} &= 0~,\\   
\int \dv x\ \{u_j(x)\overline{u_{j'}(x)}-p_j(x)\overline{p_{j'}(x)}\} &= \delta_{jj'}~.
\end{split}
\end{equation}
The reader may verify that the reduced Hamiltonian becomes diagonal in the $\alpha$-operators, provided the operator matrix equation for the basis holds:
\begin{equation}\label{BdG1}\int{\dv y\Big\{\begin{pmatrix}
 H(x,y) & -(\upsilon_N m)(x,y) \\
 \overline{(\upsilon_N m)(x,y)} & -H^T(x,y)
\end{pmatrix}\begin{pmatrix}
 u_j(y) \\ p_j(y)
\end{pmatrix}\Big\}}= E_j \begin{pmatrix}
 u_j(x) \\ p_j(x)
\end{pmatrix},\quad \|u_j\|^2 - \|p_j\|^2 = 1.
\end{equation}
In this setting, the effective Hamiltonian $\mathcal{H}_\mathrm{approx}$ reads, 
$$\mathcal{H}_\mathrm{approx} = \sum_j{E_j \alpha_j^\ast \alpha_j},\quad E_j < E_{j+1}, \quad j\in\mathbb{N}.$$
The normalization condition for $\{u_j(x),\,p_j(x)\}_{j=1}^\infty$ in \eqref{BdG1} is natural due to the constraint \eqref{eq:UP-relns}, which is a consequence of the Fock space commutation relations. Using \eqref{bformula}, Griffin finds that the functions $n^\mathrm{pair}(x,y),\,m^\mathrm{pair}(x,y)$ have the following expansion in the single particle basis $\{u_j(x),\,p_j(x)\}_{j=1}^\infty$ :
\begin{equation}\label{npairmpair}\begin{split}
n^\mathrm{pair}(x,y) = \sum_j{p_j(x)\overline{p_j(y)}},\quad
m^\mathrm{pair}(x,y) = -\sum_j{u_j(x)\overline{p_j(y)}}.
\end{split}
\end{equation}
Equations \eqref{Griffinphi}, \eqref{BdG1}, and \eqref{npairmpair} constitute the fundamental system of integro-partial-differential equations which are the subject of this work. We emphasize that \eqref{Griffinphi} and \eqref{BdG1} contain nonlinear terms which come from the pair correlation functions $m^\mathrm{pair}$ and $n^\mathrm{pair},\, \rho^\mathrm{pair}$ given in \eqref{npairmpair}. 
 
\begin{remark}
\textbf{(Reduction of Fetter, 1971 \cite{fetter72})} The above reduction can be simplified to yield a regularized version of the approximate Hamiltonian used by Fetter in \cite{fetter72}. Specifically, setting 
$m^\mathrm{pair}(x,y) = n^\mathrm{pair}(x,y) = 0$
in the above derivation results in a reduced Hamiltonian which is quadratic in the field operators $b_x,\,b_x^\ast$. In \cite{fetter72}, this approximation is accomplished without reference to two-particle correlation functions by explicitly dropping cubic and quartic terms in $b_x,\,b_x^\ast$ that result from applying the Bogoliubov approximation \eqref{Bogapprox} to $\mathcal{K}$. The Hartree-type equation for the mean-field condensate $\phi(x)$ which results is given by
\begin{equation}\label{Fetterphi}
(-\Delta_x + V_\mathrm{trap}(x))\phi(x) + (\upsilon_N\ast |\phi|^2)(x)\phi(x) = \mu\phi(x).
\end{equation}
Diagonalization of the corresponding quadratic operator results in a \textit{linear} matrix system analogous to \eqref{BdG1}. In \cite{GrillakisMargetisSorokanich}, we identify the spectrum of this linear eigenvalue problem with \textit{quasiparticle excitations} of the Bose gas. 
\end{remark}

\subsection{Connection to the method of pair excitation}
\label{subsec:prev_works} 
We now describe how the infinite nonlinear coupled system \eqref{BdG1} for basis $\{u_j(x),\,p_j(x)\}_{j=1}^\infty$ is related to a single nonlinear equation for a kernel $k(x,y)$. This is summarized in the following Proposition: 
\begin{proposition} \label{equivalencethm} 
\begin{enumerate}\item Let $\phi(x),\,m^\mathrm{pair}(x,y),\,n^\mathrm{pair}(x,y)$ be given as elements of $L^2(\mathbb{R}^3)$ and $L^2(\mathbb{R}^3\times\mathbb{R}^3)$ respectively, and let  $m(x,y),\,n(x,y)$ be defined as in \eqref{8.5}. If $k:(\phi_\perp)\to(\phi_\perp)$ is a symmetric Hilbert-Schmidt solution to the following equation
\begin{equation}\label{nonlinearRiccati}
H\circ k + k\circ H^T + (\upsilon_N m) + k\circ(\upsilon_N \overline{m})\circ k = 0,\quad \|k\|_\mathrm{op}<1,\end{equation}
then there exists a basis $\{u_j(x), p_j(x)\}_{j=1}^\infty$ for $(\phi_\perp)$, which solves \eqref{BdG1} with normalization $\|u_j\|^2 - \|p_j\|^2 = 1$. The spectrum $\{E_j\}_{j=1}^\infty$ in this case is positive and the following integral relation holds:
\begin{equation}\label{integralrelation}
\int{\dv y \{k(x,y) u_j(y)\}} = p_j(x),\quad j\in\mathbb{N}.
\end{equation}
\item Let $m^\mathrm{pair}(x,y),\, n^\mathrm{pair}(x,y)$ in $L^2(\mathbb{R}^3\times\mathbb{R}^3)$ be given and suppose the single-particle states $\phi(x),\,\{u_j(x), p_j(x)\}_{j=1}^\infty$ solve the system \eqref{Griffinphi}, and \eqref{BdG1}, where $\{u_j\}_{j=1}^\infty$is a basis of $(\phi_\perp)$, and furthermore suppose that the integral relation \eqref{integralrelation} holds for some $k:(\phi_\perp)\to(\phi_\perp)$ with $\|k\|_\mathrm{op}<1$ and all $j$. Then $k(x,y)$ solves the nonlinear operator equation \eqref{nonlinearRiccati}.  
\end{enumerate}\end{proposition}

The proof of this theorem is an adaptation of the spectral theory of J-self-adjoint operator matrices \cite{Tretter-book}; the proof is the same here, after substituting the appropriate operators, as it was in the case $\rho^\mathrm{pair},n^\mathrm{pair},m^\mathrm{pair}=0$, which was handled in \cite{GrillakisMargetisSorokanich}. We include it here for sake of completeness.
 
\begin{proof}
The matrix in equation \eqref{BdG1} is diagonalized by an operator matrix on $\phi_\perp\oplus\phi_\perp$ constructed using $k$ which solves \eqref{nonlinearRiccati}. In particular, define the operator matrices
\begin{equation*}
W := \begin{pmatrix}
\widehat\delta & \quad k \\
 \overline{k} & \quad \widehat\delta 
\end{pmatrix}
,\quad W^{-1} =\begin{pmatrix}
(\widehat\delta-k\circ\overline{k})^{-1} & \quad -k(\widehat\delta-\overline{k}\circ k)^{-1} \\
 -\overline{k}(\widehat\delta-k\circ\overline{k})^{-1} & \quad (\widehat\delta-\overline{k}\circ k)^{-1} 
\end{pmatrix}\end{equation*}
(recalling the definition $\widehat\delta(x,y) := \delta(x-y) -\phi(x)\overline{\phi(y)}$), and 
\begin{equation*} 
D := \begin{pmatrix}
 H^T + k\circ\overline{(\upsilon_N m)} & 0 \\
 0 & -H-\overline{k}\circ{(\upsilon_N m)}
\end{pmatrix}~.
\end{equation*}
A direct calculation, which uses the fact that $k$ satisfies \eqref{nonlinearRiccati},  shows that the following holds:
\begin{equation*} 
\begin{pmatrix}
 H^T + k\circ\overline{(\upsilon_N m)} & 0 \\
 0 & -H-\overline{k}\circ{(\upsilon_N m)}
\end{pmatrix}~ = W 
\circ \begin{pmatrix}
 H & -(\upsilon_N m) \\
 \overline{(\upsilon_N m)} & -H^T
\end{pmatrix}
\circ W^{-1}.
\end{equation*}
In order for both $W$ and $W^{-1}$ to be bounded, we require that $(\delta - k\circ\overline{k})^{-1}$ be a bounded operator, which is true if $\|k\|_\mathrm{op}<1$. The operator $H^T + k\circ\overline{(\upsilon_N m)}$, while not self-adjoint,  is similar to the self-adjoint operator:
\begin{equation}\label{omegasystem}
(\delta - k\circ\overline{k})^{-1/2}\circ\big(H^T+k\circ\overline{(\upsilon_N m)}\,\big)\circ(\delta - k\circ\overline{k})^{1/2},
\end{equation}
(this is a consequence of the equation \eqref{nonlinearRiccati} for $k$). The spectrum $\sigma\big(H^T + k\circ \overline{(\upsilon_N m)}\big)$ is therefore real, discrete, and bounded below (as a compact perturbation of $-\Delta + V_\mathrm{trap}(x) -\mu$), and the system of eigenvectors of \eqref{omegasystem}, denoted $\{\eta_j(x)\}_{j=1}^\infty,$ is orthogonal and complete in the space $(\phi)_\perp$. Given the solutions $\eta_j(x)$, we can construct $u_j(x),\,p_j(x)$ which solve the matrix system via 
$$u_j(x) = (\delta - k\circ\overline{k})^{-1/2}(x,\eta_j),\quad p_j(x) = k(x,u_j).$$
The spectrum of $D$ consists of two discrete and semi-bounded parts: $\sigma(H^T+k\circ\overline{(\upsilon_N m)})$ is bounded below (since $\sigma(H^T)$ is bounded below), and $-\sigma(H^T+k\circ\overline{(\upsilon_N m)})$ is bounded above. The normalization condition for $\{u_j, p_j\}$, which comes from the fact that these functions represent Bogoliubov rotations, implies that the Bogoliubov spectrum of the quadratic Hamiltonian consists of one part only, namely $\sigma(\mathcal{K}) = \sigma\big(H^T+k\circ(\upsilon_N \overline{m})\big)$. \end{proof}
 
\begin{corollary} \label{kformula} Using \eqref{npairmpair} and \eqref{integralrelation}, the following formulas hold for the density operators:
\begin{equation}\label{omegak}\begin{split}
 m^\mathrm{pair}(x,y) &= -\sum_j{u_j(x)\overline{p_j(y)}} = \Big(k\circ(\delta-\overline{k}\circ k)^{-1}\Big)(x,y) \\
n^\mathrm{pair}(x,y) &= \sum_k{p_j(x)\overline{p_j(y)}} = \Big((k\circ\overline k)\circ(\delta- \overline{k}\circ k)^{-1}\Big)(x,y).
\end{split}
\end{equation}
\end{corollary}

This Corollary shows that the terms of \eqref{nonlinearRiccati} which involve the density operators $m^\mathrm{pair},\,n^\mathrm{pair}$ can be rewritten exclusively using the kernel $k$ when system \eqref{BdG1} is satisfied. In particular, the operator kernels $H(x,y)$ and $(\upsilon_N m)(x,y)$ depend on the quantities $\phi(x),\,k(x,y)$. Motivated by Corollary \ref{kformula}, we define:
\begin{equation}\begin{split}\label{gN}
&H[\phi, k; \mu](x,y) :=\big(-\Delta_x + V_\mathrm{trap}(x)- \mu\big)\delta(x-y) \\
&\quad\quad\quad\quad\quad\quad + \overline{\phi(y)}\upsilon_N(x-y)\phi(x) + (\upsilon_N * |\phi|^2)(y)\delta(x-y)\\
&+\frac{1}{N}\upsilon_N(x-y)\Big\{\frac{k\circ\overline{k}}{\delta - k\circ\overline{k}}\Big\}(x,y) + \frac{1}{N}\int{\dv x'\upsilon_N(x-x')\Big\{\frac{k\circ\overline{k}}{\delta - k\circ\overline{k}}\Big\}(x',x')}\delta(x-y).
\end{split}
\end{equation}
We also define the operator kernel $\Theta[\phi,k](x,y)$, 
\begin{equation}\label{Theta}\begin{split}
\Theta[\phi, k](x,y) &:= \upsilon_N(x-y)\Big\{\phi(x)\phi(y) + \frac{1}{N}\big(k\circ (\delta - \overline{k}\circ k)^{-1}\big)(x,y)\Big\},
\end{split}
\end{equation}
so that the operator equation \eqref{nonlinearRiccati} for $k$ under this condition reads: 
\begin{equation}\label{nonlinearRiccati2}
H[\phi,k;\mu]\circ k + k\circ H^T[\phi,k;\mu] + \Theta[\phi,k] + k\circ \overline{\Theta[\phi,k]}\circ k =0,\quad k:(\phi)_\perp \to (\phi)_\perp.
\end{equation}
\begin{remark} \begin{enumerate}
\item{} For simplicity, we have written the equation for $k$ as it holds on the space $\phi_\perp$. This allows us to avoid including the Lagrange multiplier term $\lambda\otimes_\mathrm{s}\phi$ in the equation \eqref{nonlinearRiccati} for $k$. Also, the fraction notation $k\circ\overline{k}/(\delta - k\circ\overline{k})$ is justified here because of the commutation
$$k\circ \overline{k}\circ(\delta - k\circ\overline{k})^{-1} = (\delta - k\circ \overline{k})^{-1}\circ k\circ\overline{k}.$$
\item{} An alternative approach to the approximation of $\mathcal{K}$, which also results in the equation \eqref{nonlinearRiccati2} for the kernel $k(x,y)$, involves the transformation of $\mathcal{K}$ by the non-Hermitian pair-excitation operator, which takes the form 
$$ \mathcal{K} \mapsto \exp\big(P[k]\big)\mathcal{K}\exp\big(-P[k]\big),\quad P[k]:= -\frac{1}{2}\iint{\dv x \dv y k(x,y)b_x^\ast b_y^\ast}.$$ The transformation, and therefore $k(x,y)$, is chosen so that there are no terms in the resulting non-Hermitian operator which are proportional to the product $b_xb_y$. This criterion for determining $k$ is attributed to Wu \cite{wu58, wu61, wu98}. Non-Hermitian quantum mechanical Hamiltonians with real spectra have more recently been studied by Bender \cite{Bender2007}.
\end{enumerate}
\end{remark}
\section{Main results}
\label{sec:results}
Proposition \ref{equivalencethm} and its Corollary show that the system \eqref{Griffinphi}, \eqref{BdG1}, \eqref{npairmpair} can be recast as a coupled system for the quantities $\phi(x)$ and $k(x,y)$, which is the subject of the current analysis. We adopt a variational approach to find a unique solution to this system. 

The following function space notation will be used throughout the rest of the work. For trapping potential $V_\mathrm{trap}(x)$, we define the space 
$$L^2_V(\mathbb{R}^d) = \big\{f(x):\, \int{\dv x \{ V_\mathrm{trap}(x)|f(x)|^2\}}<\infty\big\}.$$ Function spaces on $\mathbb{R}^{d}$ (e.g., $d=3$) are denoted by lowercase gothic letters, viz., 
\begin{equation*}
    \mathfrak{h}(\mathbb{R}^d) := L^2(\mathbb{R}^d)~,\quad  \mathfrak{h}^1(\mathbb{R}^d) := H^1(\mathbb{R}^d)~, \quad \mathfrak{h}_V^1(\mathbb{R}^d) := H^1(\mathbb{R}^d)\cap L_V^2(\mathbb{R}^d)~.
\end{equation*}
We write $\mathfrak{h}$, $\mathfrak{h}^1$, $\mathfrak{h}_V^1$ for these spaces if $d=3$. As an exception to this notation, as before, we define  $(\phi_{\perp}):=
 \big\{e\in \mathfrak{h}^{1}_{V}\ \big\vert\ e\perp\phi\big\}$ where $\phi\in \mathfrak{h}^1$ is the condensate wave function. 
The space of bounded linear operators on $\mathfrak{h}$ is denoted $\mathfrak{B}(\mathfrak{h})$, with norm $\|\cdot\|_{\mathrm{op}}$. Also, the space of trace-class operators on $\mathfrak{h}$ is denoted $\mathfrak{B}_1(\mathfrak{h})$ with norm
\begin{equation*}
\|A\|_{\mathfrak{B}_1(\mathfrak{h})}=\|A\|_1= \mathrm{tr}|A|~,\quad \forall\,\,A\in\mathfrak{B}_1(\mathfrak{h})~.
\end{equation*}
Similarly, the space of Hilbert-Schmidt operators on $\mathfrak{h}$ is $\mathfrak{B}_2(\mathfrak{h})$ with norm
\begin{equation*}
\|A\|_{\mathfrak{B}_2(\mathfrak{h})}=\|A\|_2 = (\mathrm{tr}| A^\ast A|)^{1/2}~,\quad \forall\,\,A\in\mathfrak{B}_2(\mathfrak{h})~.
\end{equation*}
The space of compact operators on $\mathfrak{h}$ is $\mathfrak{B}_0(\mathfrak{h})$. 
Note the  inequalities
\begin{equation*}
\|A\|_{\mathrm{op}} \le \|A\|_{2} \le \|A\|_{1}~, 
\end{equation*}
and the inclusions
$\mathfrak{B}_1(\mathfrak{h})\subseteq\mathfrak{B}_2(\mathfrak{h})\subseteq\mathfrak{B}_0(\mathfrak{h})\subseteq\mathfrak{B}(\mathfrak{h})$. 
 
\subsection{Main theorem and variational formulation}
\begin{theorem}\label{maintheorem}
Suppose the spectrum of the single-particle operator $\big(-\Delta + V_\mathrm{trap}(x)\big)$ contains a gap between its ground state and first excited state. Also, suppose the interaction potential is given by $\upsilon_N = N^{3\beta}\upsilon_N(N^\beta x)$ for $0\le \beta\le 1$ where $\upsilon(x)$ is smooth and bounded, and $\widehat{\upsilon}\ge0$, where $\widehat{\upsilon}$ is the Fourier transform of $\upsilon$. Then for $g>0$ small enough and $0\le \beta\le 1$, there exist $\phi\in \mathfrak{h}^1_V$, with $\|\phi\|_{L^2} =1$, and $k\in\mathfrak{B}_2(\phi_\perp, \phi_\perp)$, with $\|k\|_\mathrm{op}<1$ which solve the coupled system consisting of the equation for $\phi$:

\begin{equation}\label{coupledsystem}\begin{split}
(&-\Delta_x + V_\mathrm{trap}(x))\phi(x) + (\upsilon_N\ast |\phi|^2)(x)\phi(x) \\
&+ \frac{1}{N}\big(\upsilon_N\ast \Big\{\frac{k\circ\overline{k}}{\delta - k\circ\overline{k}}\Big\}(x',x')\big)\phi(x) + \frac{1}{N}\int{dy \{\upsilon_N(x-y)\Big\{\frac{k\circ\overline{k}}{\delta - k\circ\overline{k}}\Big\}(x,y)\phi(y)\}} \\
&+ \frac{1}{N}\int{dy\{\upsilon_N(x-y)\Big\{\frac{k}{\delta - k\circ\overline{k}}\Big\}(x,y)\overline{\phi(y)}\}} = \mu\phi(x),
\end{split}\end{equation}
and the nonlinear operator equation for $k(x,y):(\phi)_\perp \to (\phi)_\perp,$
\begin{equation}\label{coupledsystem2}
H[\phi,k;\mu]\circ k + k\circ H^T[\phi,k;\mu] + \Theta[\phi,k] + k\circ \overline{\Theta[\phi,k;\mu]}\circ k =0,
\end{equation}
where the operators $H[\phi,k;\mu](x,y)$ and $\Theta[\phi,k](x,y)$ are given by formulas \eqref{gN} and \eqref{Theta} respectively.
\end{theorem}
 
The functions $\phi$ and $k$ are sought as minimizers of the total energy, $\mathcal{E}_\mathrm{tot}[\phi,k]$, which is now introduced.
 
\begin{definition}
The Hartree functional, denoted $E_\mathrm{H}$, is defined for all $\phi\in\mathfrak{h}^1_V$, via
\begin{equation}\label{GPfunctional}
E_\mathrm{H}[\phi] := \iint{\dv x\dv y\big\{\overline{\phi(x)}\big(-\Delta +V_\mathrm{trap}(x)\big)\delta(x-y)\phi(y) + \frac{1}{2}|\phi(x)|^2 \upsilon_N(x-y) |\phi(y)|^2\big\}}.
\end{equation}
Given the operators $H[\phi,k],\, \Theta[\phi,k]: \mathfrak{h}_V^1 \to \mathfrak h^1_V$ (defined by equations \eqref{gN}, \eqref{Theta}), let 
\begin{equation}
H_0[\phi,k]:= H[\phi,k;\mu] + \mu \delta(x-y),
\end{equation}
(so that $H_0[\phi,k]$ contains no contribution proportional to $\mu$). Given $\phi\in\mathfrak{h}_V^1$, the functional 
$$\mathcal{E}\big[k; H_0[\phi,k],\Theta[\phi,k]\big],\quad \mathrm{dom}(\mathcal{E}):=\big\{k\in\mathfrak{B}_2(\mathfrak{h}^1_V),\quad \|k\|_\mathrm{op}<1\big\},$$ is defined by the operator trace formula,
\begin{equation}\label{totalenergy}
\mathcal{E}[k;\, H_0, \Theta] := \mathrm{tr}\Big\{(\delta - \overline{k}\circ k)^{-1}\circ\Big(\overline{k}\circ H_0[\phi,k]\circ k + \frac{1}{2}\overline{k}\circ \Theta[\phi,k] + \frac{1}{2}\overline{\Theta[\phi,k]}\circ k \Big)\Big\}.
\end{equation}
The total energy, $\mathcal{E}_\mathrm{tot}$ is defined for all $\phi\in\mathfrak{h}_V^1$ and $\big\{k\in\mathfrak{B}_2(\mathfrak{h}_V^1),\quad \|k\|_\mathrm{op}<1\big\}$ by the sum
\begin{equation}\label{totalenergy}
\mathcal{E}_\mathrm{tot}[\phi, k; N]:=\Big\{E_\mathrm{H}[\phi]+\frac{1}{N}\mathcal{E}\big[k;\, H_0[\phi,k], \Theta[\phi,k]\big]\Big\},
\end{equation}
$$\mathrm{dom}(\mathcal{E}_\mathrm{tot}) = \{\phi\in\mathfrak{h}^1_V,\quad k\in\mathrm{dom}(\mathcal{E})\}.$$
\end{definition}
 
It is worth commenting on the manner in which the energy $\mathcal{E}[k;H_0,\Theta]$ is written in equation \eqref{totalenergy}. Specifically, it might be claimed that our use of operators $H_0[\phi,k]$ and $\Theta[\phi,k]$ obscure important nonlinear dependences of the energy on $k$. We choose to write $\mathcal{E}$ in this way in order to draw the connection to the variational method of our previous work \cite{GrillakisMargetisSorokanich}, in which the analogues of operators $H_0,\,\Theta$ did not depend on the argument $k$.

The following lemma describes how the coupled system \eqref{coupledsystem} and \eqref{coupledsystem2} can be derived from the total energy functional. The proof is direct, but rather involved. It is therefore omitted.

\begin{lemma} \label{energy} Let $H[\phi,k;\mu], \Theta[\phi, k]$ be given by \eqref{gN}, \eqref{Theta}, and let the total energy $\mathcal{E}_\mathrm{tot}$ be given by formula \eqref{totalenergy}, i.e., 
$$
\mathcal{E}_\mathrm{tot}[\phi, k; N]:=\Big\{E_\mathrm{H}[\phi]+\frac{1}{N}\mathcal{E}\big[k;\, H_0[\phi,k], \Theta[\phi,k]\big]\Big\}.
$$

Then \textbf{(i)} For fixed $k,\,\overline{k}$, critical points of the total energy
with respect to constrained variations $\overline{\phi}\in L^2(\mathbb{R}^3)$, $\|\phi\|^2=1$, i.e., points satisfying $\,\delta \mathcal{E}_\mathrm{tot}/\delta\overline{\phi}=\mu \phi\,$, where $\mu$ is the Lagrange multiplier enforcing the constraint, are solutions to the nonlinear Hartree-type equation \eqref{coupledsystem}.

\vspace{2mm}
\textbf{(ii)} For fixed $\phi$ and $\mu$ satisfying \eqref{coupledsystem}, critical points of
$$\mathcal{E}_\mathrm{tot} + \mu \, \mathrm{tr}\Big\{(\delta- \overline{k}\circ k)^{-1}\circ\Big(\overline{k}\circ k\Big)\Big\}:=E_\mathrm{H}[\phi] + \frac{1}{N}\mathcal{E}\big[k;\, H[\phi,k;\mu],\Theta[\phi,k]\big],$$
with respect to symmetric variations of $\overline{k}\in\mathfrak{B}_2(\mathfrak{h}_V)$, i.e., points satisfying 
$$\delta/\delta\overline{k} \Bigg[\mathcal{E}_\mathrm{tot} + \mu \, \mathrm{tr}\Big\{(\delta- \overline{k}\circ k)^{-1}\circ\Big(\overline{k}\circ k\Big)\Big\} \Bigg]=0,$$ are solutions to the nonlinear equation \eqref{coupledsystem2}. 
\end{lemma}
\begin{remark}
Integrating \eqref{coupledsystem} against $\overline{\phi(x)}$ gives the formula for $\mu$:
\begin{equation}\label{muformula}
\begin{split}
\mu &= \iint{\dv x\dv y \,\overline{\phi(x)}\Big(-\Delta_x +V_\mathrm{trap}(x)\Big)\delta(x-y)\phi(y) + |\phi(x)|^2\upsilon_N(x-y)|\phi(y)|^2}\\
&+\frac{1}{N}\iint{\dv x \dv y\,\Big\{\frac{k\circ\overline{k}}{\delta - \overline{k}\circ k}\Big\}(y,y)\upsilon_N(x-y)|\phi(x)|^2} \\
&+ \frac{1}{N}\iint{\dv x \dv y \Big\{\frac{k\circ\overline{k}}{\delta - \overline{k}\circ k}\Big\}(x,y)\upsilon_N(x-y)\overline{\phi(x)}\phi(y)} \\
&+\frac{1}{N}\iint{\dv x \dv y \Big\{\frac{k}{\delta - \overline{k}\circ k}\Big\}(x,y)\upsilon_N(x-y)\overline{\phi(x)}\,\,\overline{\phi(y)}}.
\end{split}
\end{equation}
\end{remark}
\subsection{Proof of Theorem \ref{maintheorem} by direct minimization}
\label{subsec:existence-summ}
The existence of a weak minimizer of $\mathcal{E}_\mathrm{tot}$, denoted by the triple $\{\phi(x),k(x,y);\mu\}$, can be demonstrated by showing that the functional $\mathcal{E}_\mathrm{tot}$ is bounded below, and extracting the weak limit of a minimizing sequence.
The proof which follows applies to the full range $0\le\beta\le 1$ of the mean-field interaction potential. However, the method of proof does not give much information about the nature of the minimizer. It also does not deal directly with the quantity $k(x,y)$, but rather a kernel denoted $\psi(x,y)$, which is an analytic function of $k(x,y)$.
 
\begin{proof}
Given $k\in\mathfrak{B}_2(\mathfrak{h}^1_V(\phi_\perp))$, with $\|k\|_\mathrm{op}<1$, we define the operator kernel $\psi(x,y)$ via a change-of-variables
\begin{equation}
\psi(x,y) := \Big(k\circ{(\delta - k\circ \overline{k})^{-1/2}}\Big)(x,y).
\end{equation}
Inverting this relationship gives 
\begin{equation}
k = \psi\circ(\delta+\overline\psi\circ\psi)^{-1/2} = (\delta + \psi\circ\overline{\psi})^{-1/2}\circ\psi,
\end{equation}
and observe that 
$\delta - \overline{k}\circ k = (\delta + \overline{\psi}\circ\psi)^{-1}.$
This constitutes a change of variables for minimization problem; the above relationship implies that we can solve the problem for $\psi$ and transform the solution to a solution $k$. The functional $\mathcal{E}_\mathrm{tot}\big[\phi,k\big]$ defined in \eqref{totalenergy}, whose minimization produces the nonlinear system for $\phi$ and $k$ is rewritten using the operator $\psi$. It can be verified that $\mathcal{E}_\mathrm{tot}[\phi,k] = \widetilde{\mathcal{E}}[\phi,\psi]$, where 
\begin{equation}\begin{split}
\widetilde{\mathcal{E}}\big[\phi,\psi\big] &:= 
N\int{\dv x\{|\nabla\phi(x)|^2+V_\mathrm{trap}(x)|\phi(x)|^2\}} \\
&+N\iint{\dv x \dv y\{\upsilon_N(x-y)\big|\phi(x)\phi(y)+\frac{1}{N}\psi\circ(\delta+\overline{\psi}\circ\psi)^{1/2}(x,y)\big|^2\}}\\
&+\frac{1}{2}\iint{\dv x\dv y\{|\nabla_x \psi(x,y)|^2 + |\nabla_y \psi(x,y)|^2 + (V_\mathrm{trap}(x)+V_\mathrm{trap}(y))|\psi(x,y)|^2\}} \\
&+\frac{1}{2}\iint{\dv x\dv y\{\upsilon_N(x-y)[\phi(x)\overline{\phi(y)}(\psi\circ\overline{\psi})(x,y)+\overline{\phi(x)}\phi(y)(\overline{\psi}\circ\psi)(x,y)]\}}  \\
&+ \iint{\dv x\dv y\{|\phi(x)|^2(\overline{\psi}\circ\psi)(y,y)+|\phi(y)|^2(\overline{\psi}\circ\psi)(x,x)\}} \\
&+\frac{1}{2N}\iint{\dv x\dv y\{\upsilon_N(x-y)\big(|(\overline{\psi}\circ\psi)(x,y)|^2+(\overline{\psi}\circ\psi)(x,x)(\overline{\psi}\circ\psi)(y,y)\big)\}}.
\end{split}\end{equation}
Recall that we assume $\widehat{\upsilon}\ge0$ as in the statement of the Theorem. This is not necessary but it simplifies the argument and makes more transparent the main idea. 

The constraint for the minimization is modified in a way that accommodates both the constraints on $\phi$ and $\psi$. Namely, we solve
$$\min \widetilde{\mathcal{E}}[\phi,\psi],\quad \mathrm{over}\quad\big\{\phi\in\mathfrak{h}^1_V,\,\psi\in\mathfrak{B}_2(\mathfrak{h}^1_V),\quad\|\phi\|_{L^2(\mathbb{R}^3)}^2 + \frac{1}{N}\|\psi\|_{L^2(\mathbb{R}^3\times\mathbb{R}^3)}^2 = 1\big\}.$$
In the new coordinate system the system can be written 
\begin{equation*}\begin{split}\frac{\delta\widetilde{\mathcal{E}}}{\delta\overline{\phi}} &= \mu\phi \\
\frac{\delta\widetilde{\mathcal{E}}}{\delta\overline{\psi}} &= \mu\psi.
\end{split}\end{equation*}
We will prove that the minimum exists and is non-trivial. Let us define the norm for $\psi(x,y)$, 
$$\|\psi\|^2_{H^1_\mathrm{trap}}:=\iint{\dv x\dv y\{|\nabla_x\psi(x,y)|^2+|\nabla_y\psi(x,y)|^2+(V_\mathrm{trap}(x)+V_\mathrm{trap}(y))|\psi(x,y)|^2\}}.$$
We know by Sobolev inequalities in $\mathbb{R}^6$, 
\begin{equation}\begin{split}
\|\psi\|_{L^3(\mathbb{R}^3\times\mathbb{R}^3)}&\le C\|\psi\|_{H^1(\mathbb{R}^3\times\mathbb{R}^3)}, \\
\|\psi\|_{L^2(\mathbb{R}^3\times\mathbb{R}^3)}&\le C \|\psi\|_{H^1_\mathrm{trap}(\mathbb{R}^3\times\mathbb{R}^3)}.
\end{split}\end{equation}
Moreover, the embedding $L^p(\mathbb{R}^3\times\mathbb{R}^3)\subset H^1_\mathrm{trap}(\mathbb{R}^3\times\mathbb{R}^3)$ is compact for $2\le p \le 3$. Defining the density $\rho^\mathrm{pair}(x):=(\overline{\psi}\circ\psi)(x,x)$, we have 
$$\|\nabla\rho^\mathrm{pair}\|_{L^2(\mathbb{R}^3)}^{3/2}\le C\|\psi\|^{3/2}_{H^1(\mathbb{R}^3\times\mathbb{R}^3)} \|\rho^\mathrm{pair}\|_{L^3(\mathbb{R}^3)}^{3/4}$$
(see Lemma \ref{rhoL3} below for a detailed proof of this). Sobolev again implies
$$\|\rho^\mathrm{pair}\|_{L^3(\mathbb{R}^3)}\le C\|\nabla\rho^\mathrm{pair}\|_{W^{1,3/2}(\mathbb{R}^3)}.$$
We also know
$$ \|\rho^\mathrm{pair}\|_{L^1(\mathbb{R}^3)} = \|\psi\|_{L^2(\mathbb{R}^3\times\mathbb{R}^3)}^2.$$
We have the inequalities:
$$ |(\overline{\psi}\circ\psi)(x,y)|\le\sqrt{\rho^\mathrm{pair}(x)\rho^\mathrm{pair}(y)}$$
and 
$$|\nabla_x(\overline{\psi}\circ\psi)(x,y)|\le \Big(\int{\dv y |\nabla_x\psi(x,y)|^2}\Big)^{1/2}(\rho^\mathrm{pair}(y))^{1/2}$$
hence:
$$ \|\overline{\psi}\circ\psi\|_{H^1(\mathbb{R}^3\times\mathbb{R}^3)}\le \|\psi\|^2_{H^1_\mathrm{trap}(\mathbb{R}^3\times\mathbb{R}^3)}$$
$$ \|\overline{\psi}\circ\psi\|_{L^3(\mathbb{R}^3\times\mathbb{R}^3)}\le \|\psi\|^2_{H^1_\mathrm{trap}(\mathbb{R}^3\times\mathbb{R}^3)}.$$
Recall the Hartree energy $E_\mathrm{H}[\phi]$, and notice that $\widetilde{\mathcal{E}}(\phi,0)=N E_\mathrm{H}[\phi]$. The trapping norm for $\phi$ is 
$$\|\phi\|_{H^1_\mathrm{trap}(\mathbb{R}^3)}^2 = \int{\dv x\{|\nabla_x\phi(x)|^2+V_\mathrm{trap}(x)|\phi(x)|^2\}}$$
Notice that if we set $\rho^\mathrm{cond}(x):=|\phi(x)|^2$ for the condensate density then 
$$\|\rho^\mathrm{cond}\|_{L^3(\mathbb{R}^3)}\le C\|\phi\|_{H^1_\mathrm{trap}(\mathbb{R}^3)}.$$
Now we have all the ingredients in order to prove the existence of a minimum. We start with a minimizing sequence $(\phi^{(j)},\psi^{(j)})$ which without loss of generality we may assume converges weakly to $(\phi_\mathrm{min},\psi_\mathrm{min})$ in $H^1_\mathrm{trap}(\mathbb{R}^3)\times H^1_\mathrm{trap}(\mathbb{R}^3\times\mathbb{R}^3)$. We know that 
$$N\|\phi_\mathrm{min}\|_{H^1_\mathrm{trap}(\mathbb{R}^3)}+\frac{1}{2}\|\psi_\mathrm{min}\|_{H^1_\mathrm{trap}(\mathbb{R}^3\times\mathbb{R}^3)}\le N E_\mathrm{H},$$
where $E_\mathrm{H}:=\min_{\|\phi\|=1} E_\mathrm{H}[\phi]$. We also know that $E_\mathrm{H}$ is finite and this fact implies that the $H^1_\mathrm{trap}$ norms of $(\phi^{(j)},\psi^{(j)})$ are bounded (for fixed $N$). Here we used the fact that the sum of lines 4 and 5 in the energy functional $\widetilde{\mathcal{E}}$ is positive. The compact embedding implies that
$$\phi^{(j)}\to\phi_\mathrm{min},\quad\mathrm{strongly}\,\,\mathrm{in}\quad L^2(\mathbb{R}^3).$$
This guarantees that the solution is non-trivial. The weak convergence is in fact strong, i.e., 
$$\phi^{(j)}\to\phi_\mathrm{min},\quad\mathrm{strongly}\,\,\mathrm{in}\quad H^1_\mathrm{trap}(\mathbb{R}^3),$$
and
$$\psi^{(j)}\to\psi_\mathrm{min},\quad\mathrm{strongly}\,\,\mathrm{in}\quad H^1_\mathrm{trap}(\mathbb{R}^3\times\mathbb{R}^3).$$
We explain this conclusion for $\psi^{(j)}$ (the conclusion for $\phi^{(j)}$ follows by a similar argument). The role of $V_\mathrm{trap}$ is to guarantee that $\|\psi_\mathrm{min}\|_{L^2}$ is dominated by $\|\psi_\mathrm{min}\|_{H^1_{\mathrm{trap}}}$. In fact, all the terms in $\widetilde{\mathcal{E}}[\phi^{(j)},\psi^{(j)}]$ converge strongly except those involving $\|\phi^{(j)}\|_{H^1_\mathrm{trap}}$ and $\|\psi^{(j)}\|_{H^1_\mathrm{trap}}$. For these terms we have, e.g., 
$$\|\psi_\mathrm{min}\|^2_{H^1_\mathrm{trap}(\mathbb{R}^3\times\mathbb{R}^3)}\le \liminf_{j}\|\psi^{(j)}\|^2_{H^1_\mathrm{trap}(\mathbb{R}^3\times\mathbb{R}^3)}.$$
If the inequality were strict it would contradict the minimization because, e.g.,
$$\widetilde{\mathcal{E}}[\phi,\psi_\mathrm{min}]<\min_\psi\widetilde{\mathcal{E}}[\phi,\psi]$$
Therefore, 
$$\|\psi_\mathrm{min}\|_{H^1_\mathrm{trap}}^2 = \liminf_j \|\psi^{(j)}\|_{H^1_\mathrm{trap}}^2$$
which implies up to a subsequence $\psi^{(j)}\to \psi_\mathrm{min}$ strongly in $H^1_\mathrm{trap}(\mathbb{R}^3\times\mathbb{R}^3)$. Thus the minimum is attained at $\psi_\mathrm{min}$ and therefore it is a weak solution.
\end{proof}
\section{Iterative Minimization} 
\label{sec:Riccati}
The starting point of the iteration scheme is the system $\{\phi_0,\,k_0=0;\, \mu_0\}$ which is the solution to \eqref{coupledsystem}, \eqref{coupledsystem2} by forcing pair, $\mathrm{pair},\,n^\mathrm{pair} \equiv 0$. In particular, the equation for $\phi_0$ is the usual Hartree equation. The equation for $k_1$ is the Riccati equation which we studied extensively in our previous work \cite{GrillakisMargetisSorokanich}. The existence result for this method is stated in Theorem \ref{theorem21}. The solution discovered by iteration corresponds to the restricted range of values $0<\beta\le 1/6$. The result nonetheless seems interesting in its own right, since the kernel found by this method will be a small perturbation of the kernel $k_1$, which has been studied before.
\begin{theorem}\label{theorem21}
Suppose the single-particle operator $\big(-\Delta + V_\mathrm{trap}(x)\big)$ contains a gap between its ground state and first excited state, and the interaction potential is given by $\upsilon_N = N^{3\beta}\upsilon_N(N^\beta x)$ for $0\le \beta\le 1/6$ where $\upsilon(x)$ is smooth and bounded. Also assume that $\widehat{\upsilon}\ge0$, where $\widehat{\upsilon}$ is the Fourier transform of $\upsilon$. Then for $g>0$ small enough, there exist functions $\phi\in \mathfrak{h}^1_V(\phi_\perp)$, with $\|\phi\|_{L^2} =1$, and $k\in\mathfrak{B}_2\big(\mathfrak{h}^1_V(\phi_\perp)\big)$, with $\|k\|_\mathrm{op}<1$ which solve the coupled system \eqref{coupledsystem}, \eqref{coupledsystem2}, for $\mu>0$.
\end{theorem}
 
The iteration provides solutions $\{\phi_n,k_n; \mu_n\}_{n=0}^\infty$ to a sequence of simpler minimization problems. It will be shown that this sequence converges strongly (in $\mathfrak{h}$ and $\mathfrak{B}(\mathfrak{h})$) to weak solutions $\{\phi, \, k;\mu\}$ of the nonlinear system. To this end, define the following shorthand for operators in this sequence: 
\begin{equation}\label{Hn}
H[\phi_n, k_n;\mu_n] := H[n],
\end{equation}
as well as  
\begin{equation}\label{Thetan}
\Theta[\phi_n, k_n]:=\Theta[n],
\end{equation}
where $H[\phi,k;\mu],\Theta[\phi,k]$ are given by the formulas \eqref{gN}, \eqref{Theta}. The iteration scheme is now summarized:
 \vspace{7mm}
\hrule
\vspace{1mm}
\textbf{Iterative Minimization for $\{\phi,\, k;\,\mu\}$:}
\vspace{1mm}
\hrule
 
\begin{itemize}
\item[] \underline{\textbf{Zero iterate:}} Set $k_0 = 0$, and solve the constrained minimization problem 
$$\min_{\|\phi\|=1}E_\mathrm{H}[\phi] ,\quad \mathrm{for}\quad \phi_0\in\mathfrak{h}.$$
This entails that $\phi_0$ satisfies $\delta E_\mathrm{H}/\delta\overline{\phi} = \mu_0 \phi$, which gives the Hartree equation 
$$\Big(-\Delta_x + V_\mathrm{trap}(x)\Big)\phi_0(x) + (\upsilon_N\ast|\phi_0|^2)(x)\phi_0(x) = \mu_0\phi_0(x).$$
For $\epsilon_0(x,y):= \big(-\Delta_x + V_\mathrm{trap}(x)\big)\delta(x-y)$, the constant $\mu_0$ is given by \eqref{muformula},
\begin{equation}
\mu_0 = \iint{\dv x\dv y \{\overline{\phi_0(x)}\epsilon_0(x,y)\phi_0(y) + |\phi_0(x)|^2\upsilon_N(x-y)|\phi_0(y)|^2\}}. \end{equation}

\item[] \underline{\textbf{($n+1$) iterate:}} Assume that the functions $\{\phi_n, \,k_n\}$ as well as the constant $\mu_n$ are given.
The operators $H[n](x,y), \,\Theta[n](x,y)$ are then defined by the formulas \eqref{Hn} and \eqref{Thetan}.
We solve the variational problem for $k = k_{n+1}$:
\begin{equation}\label{mink}
\min_{\|k\|_\mathrm{op}<1}\Big\{E_\mathrm{H}[\phi_n]+\frac{1}{N}\mathcal{E}\big[k;\,\,H[n],\,\Theta[n]\big],\quad k\in\mathfrak{B}_2(\mathfrak{h}^1_V)\Big\},
\end{equation}
and next the variational problem for $\phi = \phi_{n+1}$:
\begin{equation}\label{minphi}
\min_{\|\phi\|=1}\Big\{E_\mathrm{H}[\phi]+\frac{1}{N}\mathcal{E}\big[k_{n+1}; \,H_0[\phi,k_{n+1}], \,\Theta[\phi, k_{n+1}]\big],\quad \phi\in\mathfrak{h}^1_V\Big\}.\end{equation}
The minimization \eqref{mink} implies that $k_{n+1}:(\phi_n)_\perp \to (\phi_n)_\perp$ solves the Riccati equation
\begin{equation}\label{Riccati}
H[n]\circ k_{n+1} + k_{n+1}\circ H[n]^T + \Theta[n] + k\circ\overline{\Theta[n]}\circ k = 0,\end{equation} 
and the minimization \eqref{minphi} yields the function $\phi_{n+1}$ satisfying a Hartree-type equation of the form \eqref{coupledsystem} under the substitutions $\phi\equiv\phi_{n+1},\,k\equiv k_{n+1}$, and the constant
$\mu \equiv \mu_{n+1}$ given by \eqref{muformula}.
\end{itemize}
 
\hrule
 \vspace{7mm}
\begin{remark} (Solutions to the Riccati equation for $k_n$). Suppose the functions $\phi_n, \,k_n$ which solve \eqref{minphi}, \eqref{mink} are given, in addition to the constant $\mu_n$ and the operators $H[n],\,\Theta[n]$. We review some conclusions about minimizers of the energy functional $\mathcal{E}\big[k;\,H[n], \Theta[n]\big]$ by means of which $k_{n+1}$ is generated. Namely, if the following operator estimate holds for all $e\in(\phi_n)_\perp$:
\begin{equation}\label{opest}\Big\{H[n](e,\overline{e}) - \big|\Theta[n](\overline{e},\overline{e})\big|\Big\} \ge c \|e\|^2,\quad c>0,\quad e\in(\phi_n)_\perp,\end{equation}
then a Hilbert-Schmidt minimizer of $\mathcal{E}$, denoted $k_{n+1}\in \mathfrak{B}_2(\mathfrak{h}^1_V)$, with $\|k_{n+1}\|_\mathrm{op}<1$, exists. The proof of this statement consists in constructing an orthonormal basis $\{e_j(n+1)\}_{j=1}^\infty\subset\mathfrak{h}_V^1$ of $(\phi_n)_\perp$, and a sequence $\{z_j(n+1)\}_{j=1}^\infty\subset \ell^2$, such that
\begin{equation}\label{kexpansion}
k_{n+1}(x,y) = \sum_j{z_j(n+1) \{e_j(x)e_j(y)\}}.
\end{equation}
(We will refer to the basis as $\{e_j\}_{j=1}^\infty$ since its dependence on $n+1$ is clear from the context and does not play a role in the current argument). Plugging this expansion into the trace formula which defines $\mathcal{E}$, and setting $\partial\mathcal{E}/\partial{\overline{z_j}} = 0,$ yields two possible choices of coefficient, $z_j^{(\pm)}(n+1)$ for every $j$, where $|z_j^{+}(n+1)|<1$ and $|z_j^{-}(n+1)|>1$. The root $z_j^+(n+1)$ is the one of interest since this is the solution which guarantees $\|k\|_\mathrm{op}<1$: 
\begin{equation}\label{zj}
z_j^+(n+1) = \frac{\Theta[n](e_j, e_j)}{H[n](e_j,\overline{e_j})+\sqrt{H[n](e_j,\overline{e_j})^2 - |\Theta[n](e_j, e_j)|^2}}.
\end{equation}
This allows us to write the energy $\mathcal{E}\big[k_{n+1}; H[n],\Theta[n]\big]$ as
\begin{equation}\begin{split}
\mathcal{E}\big[k_{n+1}; H[n],\Theta[n]\big] &= \mathcal{E}\Big(\{z_j^+\},\{e_j\}\Big)\\ &=-\frac{1}{2}\sum_{j=1}^\infty{\Big\{H[n](\overline{e_j},e_j) - \sqrt{H^2[n](\overline{e_j},e_j)-|\Theta[n](\overline{e_j},\overline{e_j})|^2}\Big\}} \\
&:= -\frac{1}{2}\sum_j{\mathcal{F}(e_j)}.
\end{split}
\end{equation}
The existence of $k_{n+1}$ proceeds by maximizing $\sum_j{\mathcal{F}(e_j)}$ over all orthogonal frames $\{e_j\}$. The expression for $\mathcal{F}$ given above implies that $H[n](\overline{e},e)$ is bounded for any maximizing sequence in $(\mathrm{span}\{e_i\}_{i=1}^j)_\perp$, which allows us to extract a strongly convergent subsequence in $L^2(\mathbb{R}^3)$ to produce $e_j(x)$. Having constructed the orthonormal basis and coefficients, the proof of existence for $k_{n+1}$ is then complete.\end{remark}
 
\begin{remark} (Solutions to the nonlinear Schr\"{o}dinger-type equation for $\phi_{n+1}$) The existence of constrained minimizer $\phi_{n+1}$ of 
\begin{equation}
\min_{\|\phi\|=1}\Big\{E_\mathrm{H}[\phi]+\frac{1}{N}\mathcal{E}\big[k_{n+1}; \,H[\phi,k_{n+1}], \,\Theta[\phi, k_{n+1}]\big],\quad \phi\in\mathfrak{h}^1_V\Big\},
\end{equation}
will follow the usual proof for minimizing the Hartree functional $E_\mathrm{H}[\phi]$ (see, for example, \cite{LiebSeiringer-book}), noting that the additional terms introduced by the functional $\mathcal{E}$ are bounded when $k_{n+1}$ is fixed with $\|k_{n+1}\|_\mathrm{op}<1$. 
\end{remark}

\subsection{Preliminary lemmas}
Several preliminary results are required to carry out the proof of Theorem \ref{theorem21} via iteration, and are presented here. Lemma \ref{Tretter} is a general result on the stability of spectral gaps for relatively-bounded perturbations of self-adjoint operators. It is an extension of the work of Kato \cite{kato2012perturbation}; a concise proof can be found in the work of Cuenin and Tretter \cite{Cuenin}. Lemma \ref{rhoL3} and its Corollary prove that the kernel 
$$\rho^\mathrm{pair}[n](x):=\Big\{\frac{k_n\circ\overline{k_n}}{\delta - k_n\circ\overline{k_n}}\Big\}(x,x),$$ 
defines $(\upsilon_N\ast\rho^\mathrm{pair}[n])(x)$ as a bounded operator on $L^2(\mathbb{R}^3)$ for all $n$. The discussion which follows Lemma \ref{rhoL3} provides the existence of the sequence of functions $\{\phi_n,\,k_n\}_{n=0}^\infty$ which solve \eqref{minphi}, \eqref{mink} in the iteration scheme. To show that the induction can be carried out, it must be shown that the sequence $\Theta[n]$ remains uniformly bounded in Hilbert-Schmidt and operator norm, and that the gap condition for the operators $H[n],\,\Theta[n]$ (i.e., inequality \eqref{opest}) holds uniformly for all $n$. These estimates allow us to take strong limits of the operators $H[n],\,\Theta[n]$ as $n\to\infty$ in \eqref{minphi}, \eqref{mink} to construct solutions to the coupled system of Theorem \ref{theorem21}.

\begin{lemma} (Cuenin and Tretter, 2016 \cite{Cuenin}) \label{Tretter}
Let $T$ be a self-adjoint operator in a Hilbert space $\mathfrak{h}$ and let $A$ be bounded, i.e., $\|Ax\|  \le a\|x\| $ for all $x\in\mathfrak{h}$ with $a\ge0$. Then if $T$ has a spectral gap $(\alpha_T,\beta_T)\subset\mathbb{R}$, i.e., $\sigma(T)\cap(\alpha_T,\beta_T)=\emptyset$ with $\alpha_T,\beta_T\in\sigma(T),$ and if 
$$ 2a <\beta_T- \alpha_T,$$
then $T+A$ has a stable spectral free strip $(\alpha_{T+A},\beta_{T+A})+i\mathbb{R}\subset\mathbb{C}$, i.e.,
$$\sigma(T+sA)\cap\{z\in\mathbb{C}: \alpha_{T+A}<\mathrm{Re}(z)<\beta_{T+A}\}=\emptyset,\quad s\in[0,1],$$
with
$$\alpha_{T+A}:= \alpha_T + a,\quad\beta_{T+A}:=\beta_T- a.$$

\begin{lemma}\label{rhoL3}
Suppose that $k_{n}\in\mathfrak{B}_2(\mathfrak{h}^1_V)$  solves \eqref{mink} with $\|k_n\|_\mathrm{op}<1$, where the operators $H[n-1](x,y)$ and $\Theta[n-1](x,y)$ are given by formula \eqref{H[n]Theta[n]}. Also, suppose that the operator $\Theta[n-1](x,y)$ is Hilbert-Schmidt. Then the density 
\begin{equation}\label{rhopair}
\rho^\mathrm{pair}[n](x) := \Big(\frac{k_n\circ\overline{k_n}}{\delta - k_n\circ \overline{k_n}}\Big)(x,x),\end{equation}
satisfies $\rho^\mathrm{pair}[n](x)\in L^3(\mathbb{R}^3)\cap L^1(\mathbb{R}^3)$. In particular, the following estimate holds in Hilbert-Schmidt norm:
\begin{equation}\label{rhobounds}
\|\rho^\mathrm{pair}[n]\|_{L^3} \le C \Big\|\frac{k_n}{\delta - k_n\circ\overline{k_n}}\Big\|_\mathrm{HS}\cdot \Big(\Big\|\Theta[n-1]\Big\|_\mathrm{HS}+\mu[n-1]\Big),
\end{equation}
where $C$ is some constant independent of $n$.
\end{lemma}
\begin{proof}
Define the operator kernel $\psi(x,y)$ as in Section \ref{subsec:existence-summ} by
\begin{equation}
\psi(x,y) := \Big(\frac{k_n}{(\delta - k_n\circ \overline{k_n})^{1/2}}\Big)(x,y),
\end{equation}
so that $\rho^\mathrm{pair}[n](x) = (\psi\circ\overline\psi)(x,x)$. Then using H\"{o}lder's inequality,
\begin{equation}\begin{split}
|\nabla \rho^\mathrm{pair}[n](x)|^{(3/2)} &= \Big |\nabla_x\int{\dv z\{\psi(x,z)\overline{\psi(z,x)}\}\Big|}^{3/2} \\
&\le \Big(\int{\dv z|\nabla_x \psi(x,z)|^2}\Big)^{3/4}\Big(\int{\dv z|\psi(x,z)|^2}\Big)^{3/4}\\
&= \Big(\int{\dv z|\nabla_x \psi(x,z)|^2}\Big)^{3/4}\Big(\rho^\mathrm{pair}[n](x)\Big)^{3/4}.
\end{split}
\end{equation}
Using Holder's inequality again, ($p = 4/3,\, q = 4$) gives:
\begin{equation}\begin{split}
\int{\dv x |\nabla \rho^\mathrm{pair}[n](x)|^{3/2}} &\le  \int{\dv x\Big(\int{ \dv z|\nabla \psi(x,z)|^2}\Big)^{3/4}\Big(\rho^\mathrm{pair}[n](x)\Big)^{3/4}} \\
&\le \Big(\iint{\dv x\dv z \big|\nabla_x\psi(x,z)\big|^2}\Big)^{3/4}\Big(\int{\dv x\big(\rho^\mathrm{pair}[n](x)\big)^3}\Big)^{1/4},
\end{split}
\end{equation}
The Sobolev embedding $W^{1,\,3/2}(\mathbb{R}^3)\subseteq L^3(\mathbb{R}^3)$ gives (up to a constant)
$$\Big(\int{\dv x\rho^\mathrm{pair}[n](x)^3}\Big)^{1/3}\le \|\nabla \rho^\mathrm{pair}[n]\|_{L^{3/2}}.$$
Combining this with the previous equation, we have 
$$\Big(\int{\dv x\rho^\mathrm{pair}[n](x)^{3}}\Big)^{1/3}\le \Big(\iint{\dv x\dv z \big|\nabla_x\psi(x,z)\big|^2}\Big).$$
Now, we know that 
\begin{equation}\begin{split}
&\mathcal{E}(k_n) = \\
&\mathrm{tr}\Big\{\big(\delta - k_n\circ \overline{k_n})^{-1}\circ\Big(k_n\circ H[n-1] \circ\overline{k_n} + \frac{1}{2} k_n\circ\overline{\Theta[n-1]} + \frac{1}{2} \Theta[n-1]\circ \overline{k_n}\Big)\Big\}\le 0.
\end{split}\end{equation}
This completes the proof.
\end{proof}
 
\begin{corollary}\label{corrho}
Under the assumptions of Lemma \ref{rhoL3}, $\rho^\mathrm{pair}[n](x)\in L^2$ with 
\begin{equation}\label{rhopairwithmu}
\|\rho^\mathrm{pair}[n]\|_{L^2} \le \Big\|\frac{k_n}{\delta - k_n\circ\overline{k_n}}\Big\|_\mathrm{HS}\cdot \Big(\Big\|\Theta[n-1]\Big\|_\mathrm{HS}+\mu[n-1]\Big)^{3/4}.
\end{equation}
\end{corollary}
\begin{proof}
Interpolation gives
$$\|\rho^\mathrm{pair}[n]\|_{L^2}\le \|\rho^\mathrm{pair}[n]\|^{3/4}_{L^3} \|\rho^\mathrm{pair}[n]\|_{L^1}^{1/4},$$
and 
$$\|\rho^\mathrm{pair}[n]\|_{L^1} = \mathrm{tr}\Big(\frac{k_n\circ\overline{k_n}}{\delta - k_n\circ \overline{k_n}}\Big) \le \|k_n\|_\mathrm{op}\Big\|\frac{k_n}{\delta - k_n\circ\overline{k_n}}\Big\|_\mathrm{HS},$$
which proves the statement.
\end{proof}\end{lemma}
 
\subsection{Proof of Theorem \ref{theorem21}}
\begin{proof} We first demonstrate the existence of solutions $\{\phi_n,\,k_n;\,\mu_n\}_{n=0}^\infty$ to the iteration scheme. We also verify the gap condition for the operators $H[n],\,\Theta[n]$ at every step, showing that there exists $c>0$ independent of $n$ such that
\begin{equation} \label{gapcondition}
H[n](e,\overline{e}) - |\Theta[n](\overline{e},\overline{e})| \ge c \|e\|,\quad e\in (\phi_n)_\perp,\quad n=0,1,2,\dots.
\end{equation}
We show this inductively, beginning with $\phi_0(x)$, which is a solution to the Hartree equation:
$$\big(-\Delta_x + V_\mathrm{trap}(x)\big)\phi_0(x) + (\upsilon_N\ast|\phi_0|^2)(x)\phi_0(x) = \mu_0\phi_0(x).$$
The constant $\mu_0$ is given by  
$$\mu_0 = \iint{\dv x\dv y \{\overline{\phi_0(x)}\epsilon_0(x,y)\phi_0(y) + |\phi_0(x)|^2\upsilon_N(x-y)|\phi_0(y)|^2\}},$$ 
and the operators $H[0],\,\Theta[0]$ are defined by 
\begin{equation}\label{H[n]Theta[n]}\begin{split}
H[0] &:= \big(-\Delta_x + V_\mathrm{trap}(x)\big)\delta(x-y) +\overline{\phi_{0}(y)}\upsilon_N(x-y)\phi_{0}(x)\\
&\quad \quad  + (\upsilon_N * |\phi_{0}|^2)(y)\delta(x-y) - \mu_0,\\
\Theta[0] &:= \upsilon_N(x,y)\Big\{\phi_{0}(x)\phi_{0}(y)\Big\}.
\end{split}\end{equation}
In particular, the operator $\Theta[0]$ is Hilbert-Schmidt, with norm 
$$\|\Theta[0]\|_\mathrm{HS} \le gN^{3\beta}\|\upsilon\|_{L^\infty} \|\phi_0\|_{L^2}^2.$$
It is not required that $\Theta[0]$ must be small in this scheme, only that it is bounded (see equation \eqref{101}). Using the assumption $\widehat{\upsilon}\ge 0$ for the Fourier transform of $\upsilon$,
\begin{equation}\begin{split}
H[0](e,\overline{e})-|\Theta[0](\overline{e},\overline{e})| &\ge \int{\dv x \{e(x)\Big(-\Delta_x + V_\mathrm{trap}(x)\Big)\overline{e(x)}\}} \\
&+ \iint{\dv x\dv y\{\upsilon_N(x-y) |\phi_0(x)|^2 |e(y)|^2\}} - \mu_0 \|e\|^2.
\end{split}\end{equation}
The gap condition for $H[0],\,\Theta[0]$ follows from this inequality, because $\phi_0(x)$, as the minimizer of $E_\mathrm{H}(\phi)$, is also the ground state of the linear Hartree operator, $H_\mathrm{Hartree}$, defined by
$$H_\mathrm{Hartree}[\phi_0]:=\Big(-\Delta_x + V_\mathrm{trap}(x)\Big)\delta(x-y) + (\upsilon_N\ast|\phi_0|^2)(x)\delta(x-y).$$ 
$H_\mathrm{Hartree}$ has a discrete spectrum, with a gap between its lowest and first eigenvalues, since it is a compact perturbation of $\Big(-\Delta_x+V_\mathrm{trap}(x)\Big)$, which follows by Young's convolution inequality, and the compact Sobolev embedding $L^4(\mathbb{R}^3)\subset \mathfrak{h}^1_V$, 
\begin{equation}\label{HartreeL2bound}
\|(\upsilon_N\ast|\phi_0|^2)\|_{L^2} \le \|\upsilon_N\|_{L^1} \|~|\phi_0|^2~\|_{L^2} = \|\upsilon_N\|_{L^1} \|\phi_0\|_{L^4} \le \|\upsilon_N\|_{L^1} \|\phi_0\|_{\mathfrak{h}^1_V},
\end{equation}
i.e., 
\begin{equation}\label{HartreeL2bound2}
\|(\upsilon_N\ast|\phi_0|^2)\|_{L^2} \le \|\upsilon_N\|_{L^1} \cdot E_\mathrm{H}(\phi_0)\le g \cdot E_\mathrm{H}(\phi_0).
\end{equation}
The perturbation to $(\upsilon_N\ast|\phi_0|^2)(x)$ can be made arbitrarily small in $L^2$ by assuming that $g$ is small enough at the beginning of the induction.

For the induction step, assume that there exists $c>0$ and $C>0$ such that for $s = 0,1,2,\dots,n$, 
\begin{equation}\label{cases}
\begin{cases}
\big\|\Theta[s]\big\|_\mathrm{HS} \le C N^{3\beta},  \\
\mu[s] \le C,  \\
H[s](e,\overline{e}) - |\Theta[s](\overline{e},\overline{e})| \ge c \|e\|^2,\quad e\in(\phi_s)_\perp, 
\end{cases}
\end{equation}
We prove the inequalities of \eqref{cases} for $s=n+1$. 

The equation for $k_{n+1}$ on the space $\mathfrak{h}(\phi_\perp)$ is 
\begin{equation}\label{Ricc}
H[n]\circ k_{n+1} + k_{n+1}\circ H[n]^T + \Theta[n] + k_{n+1}\circ \overline{\Theta[n]}\circ k_{n+1} = 0,
\end{equation}
and the equation for $\phi_{n+1}$ is given by \eqref{coupledsystem} with $\phi = \phi_{n+1}$, $k = k_{n+1}$ and $\mu = \mu_{n+1}$.

To show the first inequality of \eqref{cases} for $s = n+1$, the induction assumption allows us to construct $k_{n+1}$ as a minimizer of $\mathcal{E}\big[ k; H[n],\Theta[n]\big]$, with the formula:
\begin{equation}
\Big(\frac{k_{n+1}}{\delta - k_{n+1}\circ \overline{k_{n+1}}}\Big)(x,y) = \sum_j{\Big(\frac{z_j}{1-|z_j|^2}\Big) e_j(x) e_j(y)}.
\end{equation}
Here $\{e_j(n+1)\}_{j=1}^\infty$ is an orthonormal basis of $(\phi_{n})_\perp$, which is determined as part of the construction of $k_{n+1}$. The formula for the coefficient $z_j(n+1)$ gives 
\begin{equation}
\Big|\frac{z_j(n+1)}{1-|z_j(n+1)|^2}\Big|^2 = \frac{1}{4}\Big(\frac{|\Theta[n](e_j,e_j)|^2}{H[n]^2(e_j,\overline{e_j}) - |\Theta[n](e_j,e_j)|^2}\Big) \le \frac{1}{4c^2} |\Theta[n](e_j,e_j)|^2,
\end{equation}
where the inequality follows since we assume the gap condition for the operators $H[n],\,\Theta[n]$. Thus, 
$$\|\Big(\frac{k_{n+1}}{\delta - k_{n+1}\circ \overline{k_{n+1}}}\Big)\|_\mathrm{HS} \le \frac{1}{2c}\|\Theta[n]\|_\mathrm{HS}\le\frac{C}{2c} N^{3\beta}.$$
From this (recalling that $\upsilon_N(x) = gN^{3\beta}\upsilon(N^\beta x)$ for $\beta\le 1/6$), we have
\begin{equation}\label{vnpair}
\|\frac{1}{N}\upsilon_N \Big(\frac{k_{n+1}}{\delta - k_{n+1}\circ \overline{k_{n+1}}}\Big)\|_\mathrm{HS} \le \frac{gC^2}{2c}\|\upsilon\|_{L^\infty}\Big(\frac{N^{6\beta}}{N}\Big),\end{equation} 
as well as 
\begin{equation}
\|\frac{1}{N}(\upsilon_N \Big(\frac{k_{n+1}\circ\overline{k_{n+1}}}{\delta - k_{n+1}\circ \overline{k_{n+1}}}\Big)\|_\mathrm{HS} \le \frac{gC^2}{2c}\|\upsilon\|_{L^\infty}\Big(\frac{N^{6\beta}}{N}\Big),
\end{equation}
and by Corollary \eqref{corrho} with Young's convolution inequality:
\begin{equation}\label{vnrhopair}\begin{split}
\|\frac{1}{N}(\upsilon_N\ast \rho^\mathrm{pair}[n+1])\|_{L^2} &\le \frac{gC}{2c}\frac{N^{3\beta}}{N}\cdot \Big(\Big\|\Theta[n-1]\Big\|_\mathrm{HS}+\mu[n-1]\Big)^{3/4}\\
&\le \frac{gC^2}{c} \Big(\frac{N^{6\beta}}{N}\Big).
\end{split}\end{equation}
It is now clear that for $\beta\le 1/6$ the constant $g>0$ can be chosen at the start of the induction so that 
$$\|\Theta[n+1]\|_\mathrm{HS}^2 <C N^{3\beta}.$$
This is because, e.g., 
\begin{equation}
\begin{split}
\|\Theta[n+1]\|_\mathrm{HS} &\le g N^{3\beta}\|\upsilon\|_{L^\infty}\|\phi_{n+1}\|_{L^2}^2 + \frac{1}{N} \|(\Theta^\mathrm{pair}[n+1])\|_\mathrm{HS} \\
&\le g\|\upsilon\|_{L^\infty}\Big\{N^{3\beta} + \frac{C^2}{2c}\Big\},
\end{split}
\end{equation}
where
\begin{equation}
\Theta^\mathrm{pair}[n+1](x,y):=\frac{1}{N}\upsilon_N(x-y)\Big(\frac{k_{n+1}}{\delta - k_{n+1}\circ \overline{k_{n+1}}}\Big)(x,y).
\end{equation}
We remind the reader that the inductive argument only requires $\Theta[n+1]$ to be bounded (not small), and that the particle number $N$ is large but fixed throughout the entire work. 

We use the above conclusions to show that $\mu[n+1]<C$. 
Indeed, $\phi_{n+1}$ minimizes the functional $\tilde{\mathcal{E}}[n+1](\phi)$, 
which consists of only the terms of $\mathcal{E}_\mathrm{tot}[\phi,k_{n+1}]$ which have a dependence on $\phi$, and satisfies the relation (for $n>0$),
\begin{equation}\label{93}\begin{split}
\tilde{\mathcal{E}}[n+1](\phi_{n+1}) &= \mu_{n+1} -\frac{1}{2}\iint{\dv x\dv y |\phi_{n+1}(x)|^2\upsilon_N(x-y) |\phi_{n+1}(y)|^2} \\
&- \frac{1}{2N}(\upsilon_N \Big(\frac{k_{n+1}}{\delta - k_{n+1}\circ \overline{k_{n+1}}}\Big))(\overline{\phi_{n+1}},\overline{\phi_{n+1}}).
\end{split}\end{equation}
Since $\phi_{n+1}$ is the minimizer of $\tilde{\mathcal{E}}[n+1]$, we have (recall the definition of $\rho^\mathrm{pair}[n+1](x)$ in \eqref{rhopair})
\begin{equation}\label{94}\begin{split}
\tilde{\mathcal{E}}[n+1](\phi_{n+1}) &\le \tilde{\mathcal{E}}[n+1](\phi_0) =\mu_0 -\frac{1}{2}\iint{\dv x\dv y |\phi_0(x)|^2\upsilon_N(x-y) |\phi_0(y)|^2} \\
&+\frac{1}{N}\iint{\dv x \dv y |\phi_0(x)|^2\upsilon_N(x-y)\rho^\mathrm{pair}[n+1](y)}\\
&+ \frac{1}{N}(\upsilon_N \Big(\frac{k_{n+1}\circ\overline{k_{n+1}}}{\delta - k_{n+1}\circ \overline{k_{n+1}}}\Big))(\phi_0,\overline{\phi_0}) \\
&+\frac{1}{2}\frac{1}{N}(\upsilon_N \Big(\frac{k_{n+1}}{\delta - k_{n+1}\circ \overline{k_{n+1}}}\Big))(\overline{\phi_0},\overline{\phi_0}).
\end{split}\end{equation}
Combining \eqref{93} and \eqref{94} with inequalities \eqref{vnpair}\textendash\eqref{vnrhopair},
we conclude that 
\begin{equation}\label{mu101}\begin{split}
\mu_{n+1} &\le \mu_0 + \frac{g}{2}\big(E_\mathrm{H}(\phi_0)\big)^2 + \frac{g}{2}\big(\tilde{\mathcal{E}}[n+1](\phi_{n+1})\big)^2 \\
&+\frac{2gC}{c}\|\upsilon\|_{L^\infty}+\frac{gC^2}{Nc}\mathcal{E}_\mathrm{Hartree}(\phi_0),
\end{split}\end{equation}
where we have made use of Sobolev's and Young's inequalities to write
\begin{equation}\begin{split}
\iint{\dv x\dv y |\phi_0(x)|^2\upsilon_N(x-y) |\phi_0(y)|^2} &\le E_\mathrm{H}(\phi_0)\cdot \|(\upsilon_N\ast|\phi_0|^2)\|_{L^2} \\
\iint{\dv x\dv y |\phi_{n+1}(x)|^2\upsilon_N(x-y) |\phi_{n+1}(y)|^2} &\le \tilde{\mathcal{E}}[n+1](\phi_{n+1})\cdot \|(\upsilon_N\ast|\phi_{n+1}|^2)\|_{L^2},
\end{split}\end{equation}
along with the following $L^2$ bound which is analogous to \eqref{HartreeL2bound}
$$\|(\upsilon_N\ast|\phi_{n+1}|^2)\|_{L^2} \le \|\upsilon_N\|_{L^1}\cdot \|\phi_{n+1}\|_{\mathfrak{h}^1_V} \le g\cdot \tilde{\mathcal{E}}[n+1](\phi_{n+1}),$$ 
and finally,
$$\frac{1}{N}\iint{\dv x \dv y |\phi_0(x)|^2\upsilon_N(x-y)\rho^\mathrm{pair}[n+1])(y)} \le 
\frac{g}{N} \cdot E_\mathrm{H}(\phi_0) \|\rho^\mathrm{pair}[n+1]\|_{L^2}.$$
This shows that $\mu_{n+1}$ modifies $\mu_0$ by a correction which is controlled by the constant $g$, which allows us to conclude the uniform boundedness of the constants $\mu[s]$ for $s\le n+1$.
\vspace{2mm}

Finally, we show the gap condition \eqref{gapcondition} for $s= n+1$. Observe that for any $e(x)\in (\phi_n)_\perp$,
\begin{equation}\begin{split}\label{101}
H[n+1](e,\overline{e})-|\Theta[n+1](\overline{e},\overline{e})| &\ge h[n+1](e,\overline{e})-\mu_{n+1} \\
& - \frac{1}{N}|(\upsilon_N\Big(\frac{k_{n+1}}{\delta - k_{n+1}\circ \overline{k_{n+1}}}\Big) )(\overline{e},\overline{e})|,\end{split}\end{equation}
where the operator $h[n+1]$ is defined, using $\rho^\mathrm{pair}[n+1]$ of \eqref{rhopair}, by
\begin{equation}\begin{split}
h[n+1](x,y) &:= \big(-\Delta_x + V_\mathrm{trap}(x)\big)\delta(x-y) + (\upsilon_N\ast|\phi_{n+1}|^2)(x)\delta(x-y) \\
&+\frac{1}{N}(\upsilon_N \Big(\frac{k_{n+1}\circ\overline{k_{n+1}}}{\delta - k_{n+1}\circ \overline{k_{n+1}}}\Big))(x,y) + \frac{1}{N}(\upsilon_N \ast\rho^\mathrm{pair}[n+1])(x)\delta(x-y).
\end{split}\end{equation}
Also observe that $\phi_{n+1}$ satisfies the self-adjoint operator matrix equation
\begin{equation}\label{littletheta1}
\begin{pmatrix}
h[n+1]  & \Theta^\mathrm{pair}[n+1] \\
 \overline{\Theta^\mathrm{pair}[n+1]} & h[n+1]
\end{pmatrix}\circ \begin{pmatrix}
 \phi_{n+1} \\ \overline{\phi_{n+1}}
\end{pmatrix}= \mu_{n+1} \begin{pmatrix}
 \phi_{n+1} \\ \overline{\phi_{n+1}}
\end{pmatrix},
\end{equation}
for 
\begin{equation}
\Theta^\mathrm{pair}[n+1](x,y):=\frac{1}{N}\upsilon_N(x-y)\Big(\frac{k_{n+1}}{\delta - k_{n+1}\circ \overline{k_{n+1}}}\Big)(x,y).
\end{equation}
The rest of the argument follows by showing that: \textbf{(i)} $\phi_{n+1}$ which minimizes $\tilde{\mathcal{E}}[n+1](\phi)$ is the ground state of the operator in \eqref{littletheta1}; and 
\textbf{(ii)} the higher eigenvalues of the problem \eqref{littletheta1}, which solve
\begin{equation}\label{littletheta2}
\begin{pmatrix}
h[n+1]  &  \Theta^\mathrm{pair}[n+1] \\
 \overline{\Theta^\mathrm{pair}[n+1]} & h[n+1]
\end{pmatrix}\circ \begin{pmatrix}
 e \\ \overline{e}
\end{pmatrix}= \lambda(e) \begin{pmatrix}
 e \\ \overline{e}
\end{pmatrix},\quad \lambda(e)\ge \mu_{n+1},\quad e(x)\in (\phi_{n+1})_\perp,
\end{equation}
are distinct from $\mu_{n+1}$, which has no degeneracy, i.e., 
$$\lambda(e) - \mu_{n+1} \ge \delta, \quad e\in(\phi_{n+1})_\perp, \quad \mathrm{for}\quad \delta >0.$$
The $\lambda(e)$ are given by the formula
\begin{equation}
\lambda(e) = h[n+1](e,\overline{e}) + \Theta^\mathrm{pair}[n+1](\overline{e},\overline{e}).\end{equation}
Both of these arguments follow from the fact that the matrix in equations \eqref{littletheta1}, \eqref{littletheta2} is a perturbation of the diagonal matrix $\mathrm{diag}(\epsilon_0,\,\epsilon_0)$, where $\epsilon_0$ is the single-particle Schr{\"o}dinger operator, $\epsilon_0 := -\Delta +V_\mathrm{trap}$.  
The perturbation can be written as the matrix $M'[n+1] = M_1[n+1]+M_2[n+1]$ with
\begin{equation}\begin{split}
M_1[n+1] &:= \mathrm{diag}\Big(\{(\upsilon_N\ast|\phi_{n+1}|^2)(x) + \frac{1}{N}(\upsilon_N \ast\rho^\mathrm{pair})(x)\}\delta(x-y)\\
&+\frac{1}{N}(\upsilon_N \Big(\frac{k_{n+1}\circ\overline{k_{n+1}}}{\delta - k_{n+1}\circ \overline{k_{n+1}}}\Big))(x,y)\Big),
\end{split}\end{equation}
and 
\begin{equation*}
M_2[n+1] 
:= \begin{pmatrix}
0  & \Theta^\mathrm{pair}[n+1] \\
 \overline{\Theta^\mathrm{pair}[n+1]} & 0 \end{pmatrix}.
\end{equation*}
The perturbation $M'[n+1]$ is bounded in operator norm by
\begin{equation}\label{106}
\|M'[n+1]\|_\mathrm{op}\le g\cdot \tilde{\mathcal{E}}[n+1](\phi_{n+1}) + g\|\upsilon\|_{L^\infty} C,\quad\mathrm{for}\quad C<\infty.
\end{equation}
The first term of this operator bound comes from Young's inequality (i.e., equation \eqref{mu101}). The second term in \eqref{106} meanwhile comes from \eqref{vnpair}. The important feature is that the operator bound for $M[n+1]$ can be made as small as desired by choosing $g$ small enough.

We invoke the perturbation lemma, Lemma \ref{Tretter}, in order to claim that the spectrum of the problem \eqref{littletheta2}
is a perturbation of the spectrum $\sigma(\epsilon_0)$. In particular, for $g$ small enough we can guarantee that $\mu_{n+1}$ is the ground state energy of this system, since it is the only element which is in a neighborhood of size $gC$ of the ground state of $\epsilon_0$ (where $C$ is some constant). It follows that 
$$\lambda(e) - \mu_{n+1}\ge \delta - 2 \|M'[n+1]\|_\mathrm{op},$$
for all $e\in(\phi_{n+1})_\perp$, where $\delta$ is the size of the gap for the operator $\epsilon_0.$ Since $\|M'[n+1]\|_\mathrm{op}$ can be made arbitrarily small by making $g$ small enough, we can choose the constants $g,\,c$ to be small enough so that $\lambda(e)-\mu_{n+1}>c$. This completes the proof of existence for the sequence $\{\phi_n,\,k_n;\,\mu_n\}_{n=0}^\infty$, and the uniform gap condition for the operators $H[n],\,\Theta[n]$.

We finally show that the sequence $\{\phi_n,\,k_n;\,\mu_n\}_{n=0}^\infty$ converges to the appropriate solution as described in the statement of Theorem \ref{theorem21}.
Since the gap condition holds uniformly for every step of the iteration, the sequence $\{\phi_n, k_n\}$ determined by the minimization problems \eqref{minphi} and \eqref{mink} exists, with $\|\phi_n\| = 1$ and $\|k_n\|_\mathrm{HS}<C$, $\|(\delta - k_n\circ\overline{k_n})^{-1}\|_\mathrm{op}<C $ for all $n$ and some $C$. The sequence $\{\mu_n\}$ is also uniformly bounded. 
Since we also have 
\begin{equation}\label{kSobolev}\begin{split}
\mathrm{tr}\Big\{\overline{k_n}\circ \epsilon_0 \circ k_n\Big\}&\le\Big|\mathrm{tr}\Big\{(\delta - \overline{k_n}\circ k_n)^{-1}\circ(\overline{k_n}\circ \epsilon_0\circ k_n)\Big\}\Big| \\
&\le \frac{1}{2}\Big|\mathrm{tr}\Big\{(\delta - \overline{k_n}\circ k_n )^{-1}\circ\{\overline{k_n}\circ\Theta[n-1] + \overline{\Theta[n-1]}\circ k_n\}\Big\}\Big| \\ 
&+ \mu_{n}\mathrm{tr}\Big\{(\delta - \overline{k_n}\circ k_n)^{-1}\circ(\overline{k_n}\circ k_n)\Big\}\Big|,
\end{split}
\end{equation}
the sequence $\{k_n\}$ is uniformly bounded in $\mathfrak{B}_2(\mathfrak{h}^1_V, \mathfrak{h}^1_V)$.
Because $\{\phi_n\}$ and $\{k_n\}$ are uniformly bounded in $\mathfrak{h}^1_V$, $\mathfrak{B}_2(\mathfrak{h}^1_V)$, we can find some subsequence (again denoted $\{\phi_n\}$ and $\{k_n\}$) which converges weakly in $\mathfrak{h}^1_V$, $(\mathfrak{B}_2(\mathfrak{h}^1_V)$ respectively). Since $\mathfrak{h}^1_V$ is compactly embedded in $L^2_V\cap L^4_V$, we conclude that $\phi_n$ converges strongly in $L^2_V\cap L^4_V$ to some $\phi$, and $k_n$ converges strongly in $\mathfrak{B}_2(L^2_V)$ to some $k$.

The following strong limits in $\mathfrak{B}_2(\mathfrak{h})$ therefore hold: 
\begin{equation}\begin{split}
\upsilon_N(x-y)\phi_n(x)\phi_n(y) &\to\upsilon_N(x-y)\phi(x)\phi(y), \\
\upsilon_N(x-y)\Big\{\frac{k_n}{\delta - \overline{k_n}\circ k_n}\Big\} &\to \upsilon_N(x-y)\Big\{\frac{k}{\delta - \overline{k}\circ k}\Big\}, \\
\upsilon_N(x-y)\Big\{\frac{k_n\circ\overline{k_n}}{\delta - k_n\circ\overline{k_n}}\Big\} &\to \upsilon_N(x-y)\Big\{\frac{k\circ\overline{k}}{\delta - k\circ\overline{k}}\Big\}.
\end{split}
\end{equation}
We also have the strong limits in $L^2(\mathbb{R}^3)$:
\begin{equation}\label{limits}
(\upsilon_N\ast|\phi_n|^2)(y)\delta(x-y) \to (\upsilon_N\ast |\phi|^2)(y)\delta(x-y),
\end{equation}
and
\begin{equation}\begin{split}
&\int{\dv x' \upsilon_N(x-x')\Big\{\frac{k_n\circ\overline{k_n}}{\delta - k_n\circ\overline{k_n}}\Big\}(x',x')}\delta(x-y) \\
&\to \int{\dv x' \upsilon_N(x-x')\Big\{\frac{k\circ\overline{k}}{\delta - k\circ\overline{k}}\Big\}(x',x')}\delta(x-y).
\end{split}\end{equation}
The first limit in \eqref{limits} follows by noticing 
(via Young's convolution inequality), 
$$\|(\upsilon_N\ast|\phi_n|^2)\|_{L^2}\le \|\upsilon_N\|_{L^1}\cdot\| |\phi_n|^2 \|_{L^2} = \|\upsilon_N\|_{L^1}\cdot\| \phi_n \|_{L^4},$$
and so $(\upsilon_N\ast |\phi_n|^2)(x)\to (\upsilon_N\ast|\phi|^2)(x)$ strongly in $L^2$, since $\phi_n\to\phi$ strongly in $L^4(\mathbb{R}^3)$. The second of these limits follows by the fact that $\rho^\mathrm{pair}[n](x)\in L^2(\mathbb{R}^3)$ for all $n$.

Taking $n\to\infty$ in the equations satisfied by $\phi_n, k_n$ therefore give equations \eqref{coupledsystem}, \eqref{coupledsystem2} for $\phi, k$. Finally, we must check that $k\perp \phi$. Since $k_n(\phi_n,x)=0$ by construction, we have $$\|k(\phi,x)\|_{L^2}  \le \|k\|_\mathrm{HS}\cdot\|\phi - \phi_n\|_{L^2} + \|k-k_n\|_\mathrm{HS}\cdot\|\phi_n\|_{L^2}.$$ 
By picking $n$ sufficiently large, we can bound $\|k(\phi,x)\|$ by an arbitrarily small quantity. This completes the proof. \end{proof}
\section{Conclusion} 
The result here represents an extension of the pair-excitation formalism to quadratic models for the dilute Bose gas which include the coupling of the condensate to excited states. The pair-excitation kernel introduced here is related to the diagonalization of the quadratic model in a fundamental way; the solution $k(x,y)$ can be used to  construct a basis $(u_j(x),\,p_j(x))_j$ which describes the unitary rotation of the quadratic Hamiltonian. 

The fact that the mathematical formalism of \cite{GrillakisMargetisSorokanich} can be adapted in the presence of more complicated mean-field couplings perhaps lends credit to understanding pair-excitation as a fundamental physical mechanism. The model presented here is moreover amenable to further generalization. In \cite{Griffin1996}, Griffin describes the contribution of thermal mean-field effects through finite-temperature averages, i.e., replacing the ground state expectations with thermal expectations:
$\langle b_x b_y\rangle_{T>0}$.
Such expectations involve the Bose-Einstein statistics for indistinguishable bosons. A rigorous treatment of this system would necessitate, along with the basis $\{u_j, p_j\}$, a procedure to determine the thermal mean-field averages given by 
\begin{equation}\label{bosedist}
\langle \alpha_j^\ast \alpha_k \rangle = N_0(E_j)\delta_{jk}\, ,\quad N_0(E_j):= \Big(\frac{1}{e^{\beta(T) E_j}-1}\Big),\quad \beta(T):=1/k_BT, \quad T\ge0.\end{equation}
The results presented in this work thus constitute the T=0 case.
\section*{Acknowledgements} The author would like to thank Professor Manoussos Grillakis for his guidance, useful discussions, and for his help on the proof of Theorem \ref{maintheorem}. The author would also like to acknowledge Professor Dionisios Margetis for inspiring the study of this system.
\bibliographystyle{amsplain}
\bibliography{Biblio}

\end{document}